\documentclass[amssymb,amsmath]{revtex4}

\def\t{\tau}
\def\d{\delta}
\def\D{\Delta}
\def\m{\mu}
\def\ket{\rangle}
\def\bra{\langle}
\def\a{\alpha}
\def\l{\lambda}
\def\round{\partial}
\def\g{\gamma}
\def\y{\eta}

\usepackage{amsmath,amssymb}
\usepackage{graphicx}

\begin{document}

\title{Size-frequency distribution and the large deviation function for frequency in simple models of earthquakes: a
scaling approach}

\author{Tetsuya Mitsudo$^{1,2}$}
\email{mitsudo@sat.t.u-tokyo.ac.jp}
\affiliation{$^1$FIRST, Aihara Innovative Mathematical Modelling Project,
JST, 4-6-1 Komaba, Meguro, Tokyo 153-8505, Japan. \\
$^2$Institute of Industrial Science, the University of Tokyo,
4-6-1 Komaba, Meguro, Tokyo 153-8505, Japan.}
\author{Naoyuki Kato$^3$}
\affiliation{
$^3$Earthquake Research Institute, the University of Tokyo,
1-1-1 Yayoi, Bunkyo, Tokyo, Japan.}
\date{\today}

\begin{abstract}
Fluctuations in the occurrence of large, disastrous earthquakes are important for the study of deviations from the regular behavior of
 earthquakes.
In this study, to assist in our understanding of the
 irregular behavior of earthquake occurrences, we calculate the large deviation function for the frequency of
 earthquakes.
We study the temporal sequence of the largest earthquakes in simple
 one-dimensional forest-fire models
in which the fluctuations in the loading and fracture processes are taken into
 consideration.
We introduce four different models with fixed trigger sites that represent the
 points from which ruptures propagate.
The size-frequency distributions and scaled large deviation functions for the
 frequency of the largest earthquakes in the system are calculated and
 their behaviors are classified.
The calculated large deviation functions are compared
 with those of the homogeneous Poisson process and of the one-site
 forest-fire model.
We find that the large deviation function largely depends on the model
 parameters and the fixed trigger sites, and in most cases, the large deviation function deviates from that of the homogeneous Poisson process.
The relation between the size-frequency distribution and the large deviation function for the frequency is discussed.
\end{abstract}

\maketitle

\section{Introduction}

Statistical indices for characterizing earthquakes are important for
understanding the mechanism of earthquakes. The Gutenberg-Richter (GR)
law\cite{GR} for the magnitudes of earthquakes is well established in
seismology; it states that the size-frequency distribution of earthquakes obeys a
power law.
It is known that the b-value, the exponent of the power law,
depends on both time and space \cite{Mai}. However, the functional form of the tail of the
distribution, which corresponds to large-magnitude earthquakes,
is uncertain because
we have insufficient data. 
The ability
to estimate the frequency or probability of rare events, such as
disastrous earthquakes, is also important for hazard
assessment.

To investigate the frequency of large earthquakes, we adopted a
large deviation function (LDF) \cite{Tou,Sor}, which is related to the
probability of the
rare events that constitute the tail of the probability
distribution. The
LDF is universal in the sense that it is an asymptotic form whenever the
number of elements is large.
Recently, in the field of 
nonequilibrium statistical physics, a lot of attention has been paid to the LDF for current \cite{Der}, which is expected
to serve as a thermodynamic function \cite{Nem}.
An LDF is also used in the analysis of activity in glassy systems,
at the transition between an active state and an inactive state \cite{Gar07}.
Besides the LDFs for current and activity, the LDF for frequency
has also been of interest recently for counting processes, such as photon counting \cite{Gar10,Bud11,Li11}.
Budini \cite{Bud11} studied the thermodynamic framework of a point process
by using the LDF for temporal frequency.
An earthquake sequence can be regarded as a point process, such as in the well-known epidemic-type aftershock sequence (ETAS) model \cite{Oga}.
A thermodynamic approach can also be applied to an earthquake point
process.
Recently, Monte Carlo methods have been introduced to
obtain the LDF of a system described by master equations in
discrete time \cite{Gia06} and in continuous time \cite{Lec07,Gia11}.
These methods enable us to calculate the LDF for a complex system of
earthquakes.

Because the recurrence time of large earthquakes is longer than several
tens of years, we do not have enough data, and thus
observational studies of the frequency of large earthquakes are limited.
In order to overcome this, we can generate enough data by simulating earthquakes.
In this paper, we study the LDF for the frequency of simulated earthquakes
in a one-dimensional forest-fire model, which can be understood as a
minimalist model for earthquakes.
Originally, this model was introduced to simulate forest fires
\cite{Bak90}.
Drossel and Schwabl \cite{Dro92} represented the forest fire in the model
with four processes:
planting of trees, ignition, propagation of the fire, and extinguishing of the fire.
To separate the timescale of the planting process from those of the
latter three, an effective forest-fire model was introduced
\cite{He93,Pac93}.
The effective model reduces the last three processes into a single
process, the vanishing of a cluster of trees.

Besides recent applications to real forest fires \cite{Yod11}, 
forest-fire models have been applied to the simulation of earthquakes \cite{Tur99}.
When used as an earthquake model, the loading on a fault corresponds to the planting of a
tree, and triggering an earthquake corresponds to igniting a fire.
The idea of representing an earthquake with a randomly expanded cluster
was introduced by Otsuka \cite{Ots}.
Newman and Turcotte \cite{New} studied the cycles of large earthquakes, which were
represented as percolated clusters in a 2D forest-fire model.
To establish a minimalist model of earthquakes and to estimate the
predictability of earthquakes, V\'azquez-Prada et al. \cite{Pra} arrived
at a model
similar to the 1D effective forest-fire model.
Recently, heterogeneous configurations of ignition (trigger) sites were
introduced by Tejedor et al. \cite{Tej} into 2D forest-fire models to
represent the variation in faults.
Tejedor et al. found that the size-frequency distributions of the simulated
earthquakes could be expressed by power laws.
An asperity region, where large earthquakes tend to occur repeatedly
\cite{Kaw}, can be simulated by introducing a heterogeneity.
Tejedor et al. classified the behaviors of the size-frequency distribution and
found that they depend on the configuration of the
trigger sites, and the region between the trigger sites seems to
correspond to the asperity region.

In this study, we adopted four forest-fire models with
different numbers of trigger sites as simple models of earthquakes.
Except for those that occur in the deep part of subducting
slabs (deep-focus earthquakes),
earthquake ruptures are confined to the earth's crust and are shallower than about
$50$ km, although the fault length may extend to more than $1000$ km, such
as was the case with the 2004 Sumatra-Andaman earthquake of magnitude 9.3 \cite{Amo}.
This indicates that the 1D models may be useful for considering the behavior
of large earthquakes that cut through the entire depth of the fault.
We obtained the size-frequency distribution and numerically calculated the
LDFs for the frequency of system-size earthquakes, which are
earthquakes whose size is
characterized by the system size of the model and are thus the largest
in the system. 
For system-size earthquakes, we classified into ``phases'' the behaviors of the size-frequency distribution and the
LDF.

First, we introduce the four models used in this study and present their
master equations.
We then give a brief introduction to the LDF used in this study and to how it was calculated.
Next, we show the size-frequency distributions of each of the
four models and present a table of their phases.
Similarly, we calculated the LDFs for the frequency of the system-size earthquakes,
scaled them by the frequency that minimizes the LDF, and thus present a
table of phases of the LDFs.
The distributions of the time intervals between successive system-size
earthquakes were examined in order to better understand 
the phases of the LDFs.
We discuss the relationship between the phase of the size-frequency
distribution and that of the LDF, and we present our conclusions.

\section{Model and Large Deviation Function}

\subsection{One-dimensional forest-fire models}

We studied 1D forest-fire models on a lattice of length $L$.
To take into account the heterogeneous nature of faults, we introduced
four models, M1, M2, M3, and MA, as shown in Fig.~1.
M1 has a trigger site only at the left edge, M2 can have triggers at
both the edges, M3 can have triggers at both the edges and at the site
$m$ ($2\leq m\leq L-1$), and MA can have triggers at any of the sites.
MA is a common effective forest-fire model, while
M1, M2, and M3 can be used to represent heterogeneous faults.
In M1, M2, and M3, earthquakes of various sizes can nucleate at the
trigger sites, and so the other sites are broken only by large earthquakes
that are nucleated at the trigger sites.

A steady-state solution of a master equation depends on the update rule
\cite{Sch}, and for the models introduced here, we adopted a random update rule.
For example, the results of a
random update rule are consistent with steady-state solutions
\cite{Der93,Sas98} of the master equation in continuous time for an
asymmetric simple exclusion process \cite{Raj98} when the time-step
adopted in the simulation is infinitesimally small.
For the effective forest-fire model, a formulation of the master equation in
continuous time is available \cite{Hon}, and this enables us to numerically
calculate the LDF.
In contrast, for the original forest-fire model, the calculation method for the LDF is not known.
Thus we adopted the random update rule in the models M1, M2, M3,
and MA, and set the time step to be sufficiently small.
The time-step is, ideally, infinitesimally small in order to generate
continuous time results, but we needed a practical time step in order to perform the model
simulations.
Below, we will show that the simulation with this random update rule can
produce a result that is consistent with other methods for determining
the fluctuations in the frequency of system-size earthquakes.

The procedure for the simulation is as follows. First, a site is chosen at
random.
If the site is empty, the site is loaded by the probability $p\D t$,
where $\D t$ is a small time interval.
If the site is loaded and the chosen site is a trigger site, an
earthquake is triggered with the probability $f\D t$.
Subsequently, a site is chosen, again at random.
We define a unit time step as $L/\D t$ loops of this procedure. 
When an earthquake is triggered, 
the loaded neighboring sites are also triggered; this continues until an unloaded site is encountered.
The probabilities of the loading and triggering processes describe the
fluctuations in the dynamics of the loading and rupturing processes.

\begin{figure}
\begin{center}
\includegraphics[scale=0.5]{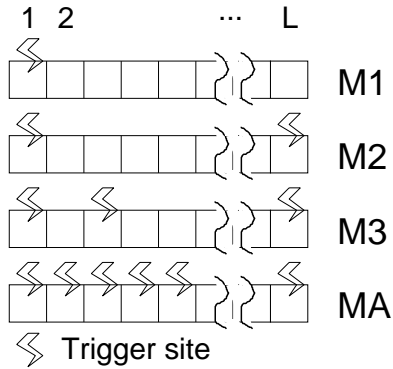}
\caption{Description of models M1, M2, M3, and MA, with lattices of length
 $L$. Trigger sites are represented by hollow zigzags (``lightning marks'').}
\end{center}
\end{figure}

The state of the $j$th site ($j=1,\cdots,L$) is described by the
occupation number $\t_j$, which is $1$ when the site is loaded and $0$
otherwise.
The loading of stress on the $j$th site is expressed as a transition from
$\t_j'=0$ to $\t_j=1$, and the release of stress (an earthquake) is expressed as a transition
from $\t_j'=1$ to $\t_j=0$, where $'$ represents the configuration before
the transition.
Suppose the sites from $j+1$ to
$j+s$ are loaded, and the sites $j$ and $j+s+1$ are empty. 
When an earthquake is triggered at one of the loaded sites,
it will be of size $s$, and it is expressed by the transition
from $\{\t_{j}',\cdots,\t_{j+s+1}'\}=\{0,1,\cdots,1,0\}$ to 
$\{\t_{j},\cdots,\t_{j+s+1}\}=\{0,\cdots,0\}$.

For simplicity, we write the configuration of the system as
$C=\{\t_1,\cdots,\t_L\}$.
We introduce the probability $P(C;t)$ that the system is in the configuration
$C$ at time $t$.
The master equation is written as
\begin{equation}
 \frac{{\rm d} P(C;t)}{{\rm d} t} = \sum_{C'\neq
  C}[W(CC')P(C';t)-W(C'C)P(C;t)],
\end{equation}
where $W(CC')$ is the transition rate from $C'$ to $C$.
The summation $\sum_{C'\neq C}$ represents the sum over all
configurations of $C'$ except for $C$.
The exact forms of $W(CC')$ are given in the appendix. 

\subsection{Large deviation function}

The mean frequency $x(s)$ of earthquakes of size $s$ per unit time $t$ is
written as
\begin{equation}
\label{meanxs}
 x(s)=\frac{N(s)}{t}.
\end{equation}
Here, $N(s)$ is the number of earthquakes of size $s$ for elapsed
time $t$.
We may write $N(s)$ as
\begin{equation}
 N(s)=\sum_{i=1}^{N_T} X(s;C_{i+1},C_{i}),
\end{equation}
where $X(s;C_{i+1},C_{i})$ is $1$ when an earthquake of size $s$ occurs
and is $0$ otherwise, when the configuration changes from $C_{i}$ to
$C_{i+1}$,
and $N_T$ is the total number of configuration changes.
The probability that the mean value $x(s)$ is equal to the frequency
$x$ is written as $P(x)$.
$P(x)$ is asymptotically written as
\begin{equation}
 P(x) \sim e^{-t\phi(x)}
\end{equation}
for large $t$, where the function $\phi(x)$ is called a large deviation
function (LDF), and it has
a minimum at $x_m(s)$, where $\phi(x_m(s))=0$ and
$\phi(x)\geq0$.
In the limit as $t\to\infty$, the statistical mean $x(s)$ converges to
$x_m(s)$.
For i.i.d. random variables, the central limit theorem can be used to
estimate small fluctuations around a mean value, but for earthquakes, $X(s;C_{2},C_{1}),\cdots,X(s;C_{N_T},C_{N_T-1})$ are not
i.i.d. random variables.
In the case of earthquakes, we note that the LDF is also significant for large
fluctuations, because of the non-i.i.d. nature.

A large deviation function $\phi(x)$ has a corresponding generating function
$\m(\l)$, where $\l$ is the conjugate variable of $x$.
The generating function $\m(\l)$ is defined as
\begin{equation}
 e^{\m(\l)t}=\bra e^{\l x t} \ket \sim \int e^{(\l x-\phi(x))t} dx,
\end{equation}
and $\phi(x)$ and $\m(\l)$ are related by the Legendre transform as
\begin{equation}
\label{legendre}
 \phi(x)=\max_{\l}[x\l-\m(\l)].
\end{equation}

For example, the LDF for the frequency of events in
a point process that obeys a homogeneous Poisson process is 
\begin{equation}
\label{ldfpoisson}
 \phi_P(x)=x \log(\frac{x}{\a})-x+\a,
\end{equation}
and the corresponding generating function is 
\begin{equation}
 \m_P(\l)=\a(e^{\l}-1),
\end{equation}
where $\a$ is the rate at which the events occur and the suffix $P$ represents
the Poisson process.

To determine the generating function $\m(\l)$, we
introduce a modified master equation written as
\begin{eqnarray}
  \frac{{\rm d}}{{\rm d} t}P^{\l}(C;t)=
\sum_{C'\ne C}[W^{\l}(CC')P^{\l}(C';t)- W(C'C)P^{\l}(C;t)],
\label{masterL}
\end{eqnarray}
where $P^{\l}(C;t)$ satisfies the differential equation (\ref{masterL})
under the initial condition $P^{\l}(C_0;t_0)=P(C_0;t_0)$ with an initial
configuration $C_0$, and $W^{\l}(CC')=W(CC')e^{\l X(s;C,C')}$.
This modified master equation is expressed by a matrix called a
modified transition matrix, and by a vector in which $P^{\l}(C;t)$ is aligned
for all $C$.
The largest eigenvalue of the matrix is asymptotically equal to $\m(\l)$
when $t$ is large.
We show here the calculation of the one-site forest-fire model to illustrate
the derivation of $\m(\l)$.
For the one-site model, the transition rate from $C'=\{0\}$ to $C=\{1\}$
is written as $W(\{1\}\{0\})$, which is equal to the loading rate $p$, and the
transition rate from $C'=\{1\}$ to $C=\{0\}$ is written as
$W(\{0\}\{1\})$, which is equal to the triggering rate $f$.
The master equation of the one-site forest-fire model is written in a
matrix form as
\begin{equation}
 \frac{{\rm d}}{{\rm d} t}\left(\begin{array}{c} P(\{0\};t) \\ P(\{1\};t)
				\end{array}\right)
=\mathsf{W}\left(\begin{array}{c} P(\{0\};t) \\ P(\{1\};t)
				\end{array}\right),
\end{equation}
with the transition matrix
$\mathsf{W}$ written as
\begin{equation}
\mathsf{W}=
 \left(\begin{array}{cc}
-W(\{1\}\{0\}) & W(\{0\}\{1\}) \\ W(\{1\}\{0\}) & -W(\{0\}\{1\}) 
\end{array}\right)
=
 \left(\begin{array}{cc}
-p & f \\ p & -f 
\end{array}\right).
\end{equation}
The modified transition matrix $\mathsf{W}^{\l}$ is written as
\begin{equation}
\mathsf{W}^{\l}=
 \left(\begin{array}{cc}
-W(\{1\}\{0\}) & W(\{0\}\{1\})e^{\l X(1;\{0\},\{1\})} \\ W(\{1\}\{0\}) & -W(\{0\}\{1\})
\end{array}\right)
=
 \left(\begin{array}{cc}
-p & fe^{\l} \\ p & -f 
\end{array}\right),
\end{equation}
with the modified master equation 
\begin{equation}
 \frac{{\rm d}}{{\rm d} t}\left(\begin{array}{c} P^{\l}(\{0\};t) \\ P^{\l}(\{1\};t)
				\end{array}\right)
=\mathsf{W}^{\l}\left(\begin{array}{c} P^{\l}(\{0\};t) \\ P^{\l}(\{1\};t)
				\end{array}\right).
\end{equation}
After calculating the largest eigenvalue of this modified matrix, the
generating function of the one-site system $\m_1(\l)$ can be given as
\begin{equation}
 \m_1(\l)=\frac{1}{2}\left[-p-f+\sqrt{(p-f)^2+4pfe^{\l}}\right].
\end{equation}
The LDF is calculated from the generating function using the Legendre
transform relation (\ref{legendre}).
By determining $\l^*$ that satisfies 
$\frac{\round}{\round \l}(x\l-\m(\l))|_{\l=\l^*}=0$, the
LDF for the frequency of earthquakes of the one-site system $\phi_1(x)$ is
obtained as
\begin{eqnarray}
\label{onesite}
 \phi_1(x) &=&
  x\log{\left[\frac{x}{pf}(2x+\sqrt{4x^2+(p-f)^2})\right]}+\frac{p+f}{2}
  \nonumber \\
& & -\frac{1}{2}\sqrt{(p-f)^2+8x^2+4x\sqrt{4x^2+(p-f)^2}}.
\end{eqnarray}
The mean frequency of earthquakes in the one-site system is
$x_{1m}(1)=\frac{pf}{p+f}$. 
This is the matrix method for obtaining the LDF.

In the present study, in order to obtain the generating
functions, the largest eigenvalue of the modified transition
matrix is calculated numerically for $L<14$.
A cloning Monte Carlo method can also be used to calculate the
generating function \cite{Lec07} for larger
systems, in which the matrix method cannot be used because the
requirements for memory size and computation time would be unrealistic.
However, when this is possible, the generating function can be calculated more accurately by the matrix
method than by the cloning Monte Carlo method.
In a normal Monte Carlo simulation, each sample evolves
independently and all the samples are usable; here, samples are 
called clones.
In the cloning method, we create the desired amount of deviation by duplicating or retaining the 
necessary clones and discarding the unnecessary ones.
Details of the cloning Monte Carlo method used in this study are given elsewhere \cite{Lec07,Gia11,Mit11}, so we will just give an
outline of the method here.

For a given initial state, we prepare $N_C$ clones,
each with its own time $t_k$ $(k=1,\cdots,N_C)$.
In each transition step, the earliest clone, here labeled $A$,
is chosen, and it evolves to a new configuration at a probability proportional
to a modified transition rate.
The initial time of $t_k$ is $0$ for all $k$, and in the first sweep, the
clones are chosen in order with respect to $k$.
After determining the transition, two values are calculated:
 $\Delta t_{A}$ and $y$, where 
$\Delta t_{A}$ is the time that has elapsed since the
previous transition, and $t_{A}$ is renewed as 
 $t_{A}\rightarrow t_{A}+\Delta t_{A}$. 
Here, $\Delta t_A$ is given by an exponential random number 
with the mean time interval $1/\sum_{C'\neq C}W(C'C)$,
$y$ is the number of clones to be copied or pruned,
$y$ is given by $y=[Y+\xi]-1$, where $\xi$ is a uniform random number
in $[0,1)$, and 
$Y=\exp(\Delta t_A(\sum_{C'\neq C}[W^{\l}(C'C)-W(C'C))])$.
During the process of copying and pruning, the number of clones $N_C$ is
kept constant. This is done by replacing a random clone when one is copied and by 
adding a copy of a random clone when one is pruned.
The generating function is calculated iteratively from the values $N_C$
and $y$.

The cloning Monte Carlo method and the matrix method can
calculate the LDF for a wider range of $x$ than can the normal Monte Carlo
method, which was the method we used to simulate the forest-fire model.
Values of $P(x)$ and $\phi(x)$ for the system-size earthquakes
that were calculated by the three different methods are shown in Fig.~2 for the
model M2 with $p=1.0, f=0.1$, and $L=12$. 
For the normal Monte Carlo for large $t$, $\phi(x)$ is approximated by
$\phi_{nM}(x)=\frac{1}{t}\log P_{nM}(x)$, and $P(x)$ of the cloning Monte Carlo
and matrix method are approximately given by
$P_{cl}(x)=e^{-t\phi_{cl}(x)}$.
For the normal Monte Carlo, $2^{14}$ and $2^{17}$ time steps and
$2^{14}$ ensemble members were simulated, and $P_{nM}(x)$ was obtained by
making a histogram of $x$.
$2^{11}$ clones were used in the cloning Monte Carlo.
For $P_{nM}(x)$ and $\phi_{nM}(x)$ with $2^{14}$ time steps, the value 
of $x$ exists in the range $0.11\leq x\leq0.125$, as shown in Fig.~2, and 
the simulation results for the other values of $x$ were not
within the ensemble of the present calculations.
This finite limiting range becomes narrower as the number of time steps increases in 
the normal Monte Carlo. The range of the $2^{17}$ time steps is
narrower than that of the $2^{14}$ time steps.
\begin{figure}
\begin{center}
\includegraphics[scale=0.3]{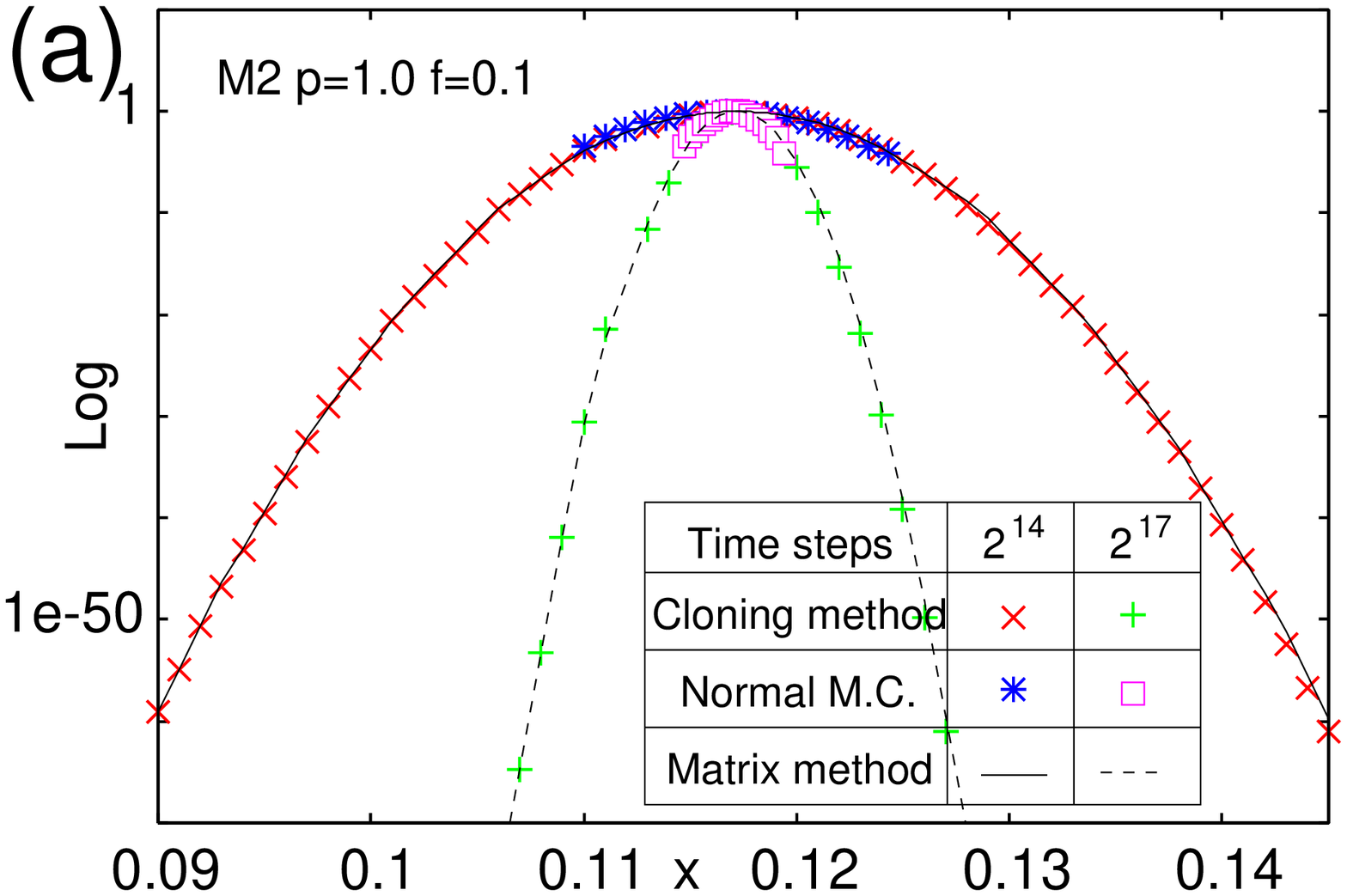}
\includegraphics[scale=0.3]{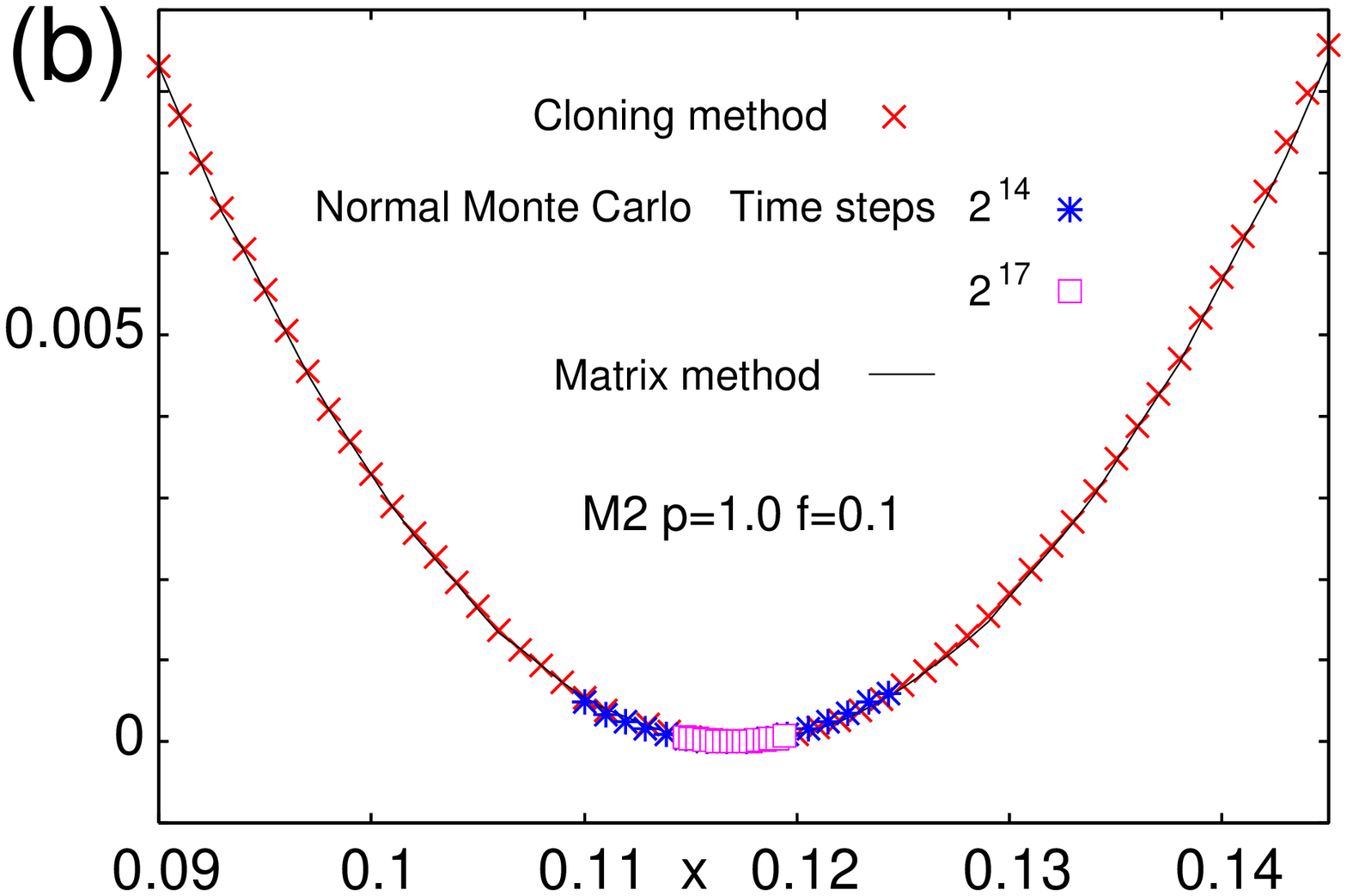}
\caption{(a) Comparison of $P(x)$ obtained by the
 normal Monte Carlo (M.C.) method, the cloning Monte Carlo method, and the
 matrix method, with different time steps. (b) Comparison of $\phi(x)$
 obtained by the same methods as in (a).}
\end{center}
\end{figure}
Figure~2 clearly shows that the results obtained from the three methods
are in close agreement.
This indicates that the methods are giving sufficiently accurate and consistent
 values.

\section{Simulation Results}

\subsection{Size-frequency distributions}

In this study, the loading rate $p$ is fixed at $1.0$, which corresponds
 to the rescaling of time by $p$.
Figure~3 shows the size-frequency distributions of simulated
earthquakes for the models M1, M2, M3, and MA, with $L=128$.
The trigger sites for M3 were located at sites $1, 32$, and $128$.
The triggering rate $f$ varies as $1.0, 0.1, 0.01$, and $0.001$.
We took $2^{27}$ time steps and recorded the number of earthquakes by
size.
The solid line in each graph of Fig.~3 denotes the function $1/s$, where $s$ is
the earthquake size.
The frequencies of the system-size earthquakes are
significantly large, with peaks at $s=L$, except for the case of MA with
$f=0.1$ or $1.0$.

\begin{figure}
\begin{center}
\includegraphics[scale=0.35]{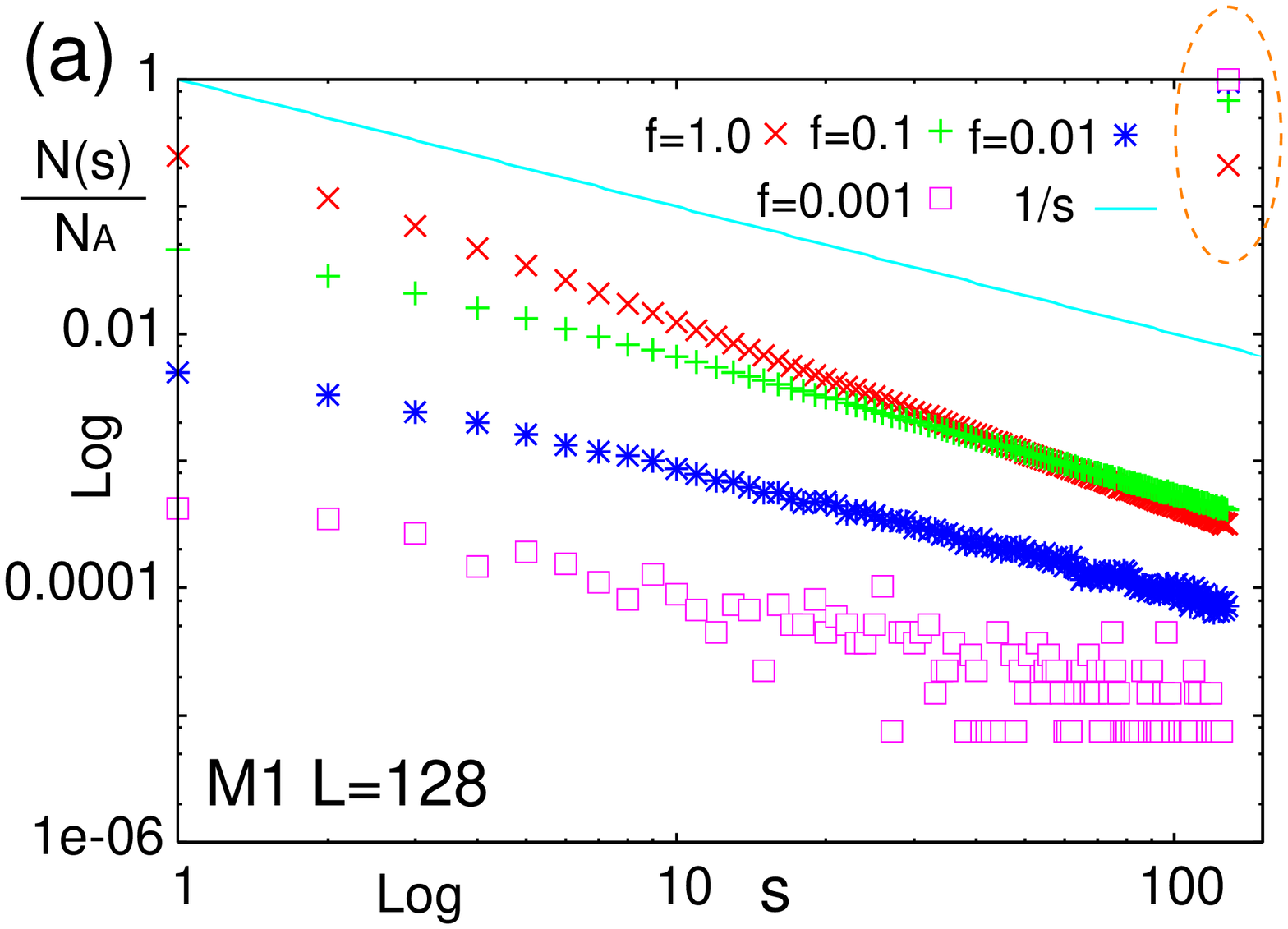}
\includegraphics[scale=0.35]{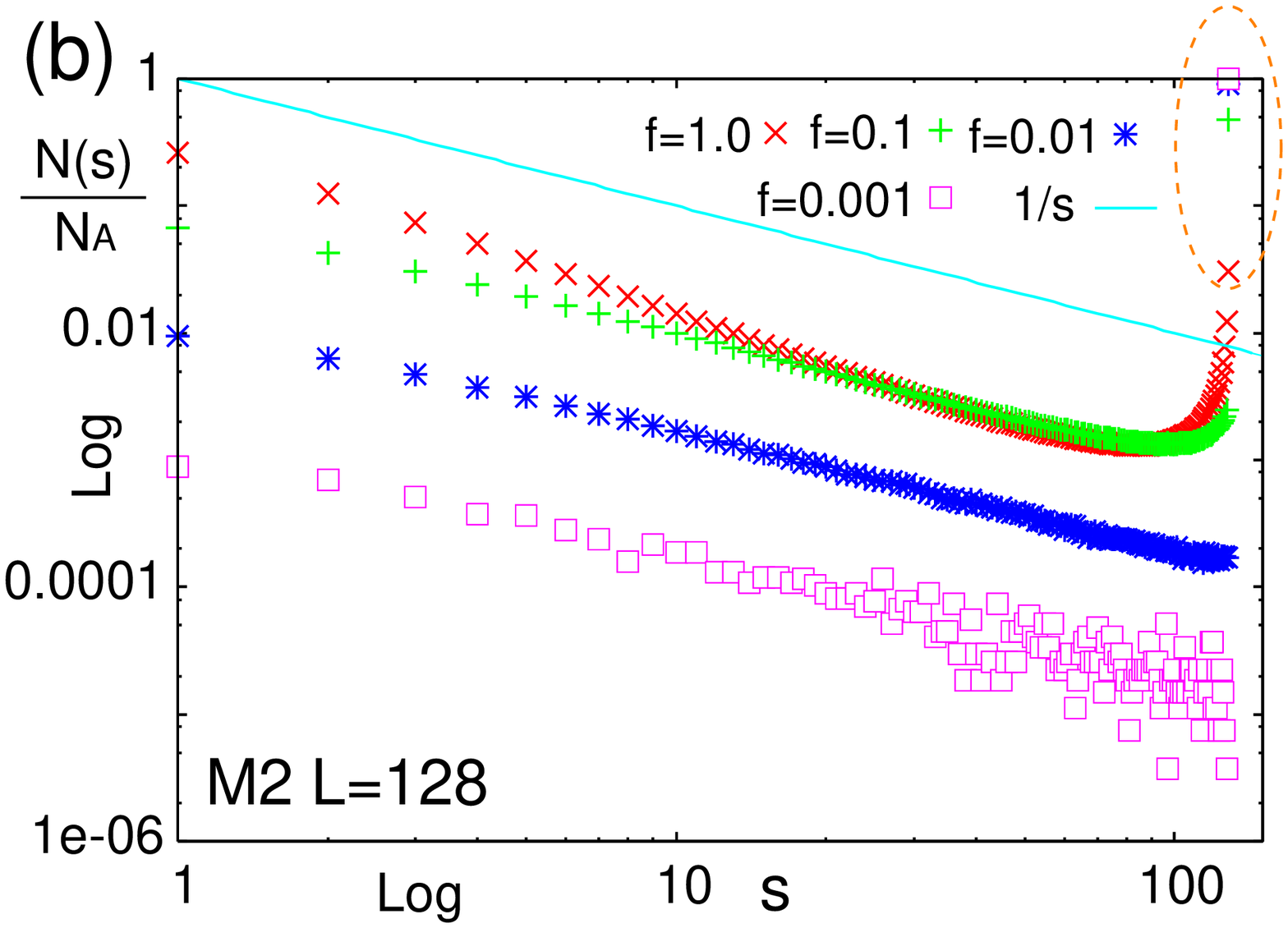}
\includegraphics[scale=0.35]{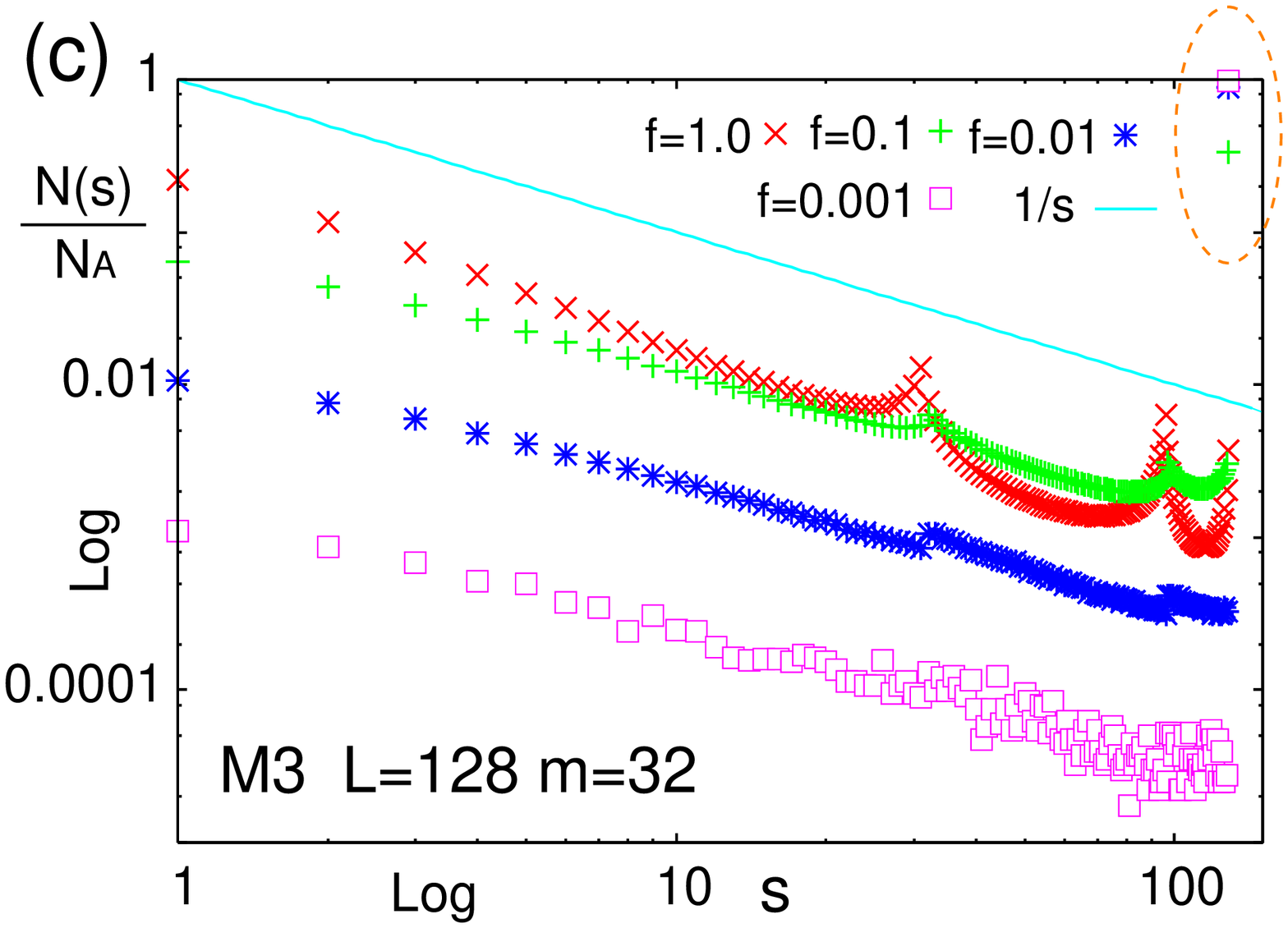}
\includegraphics[scale=0.35]{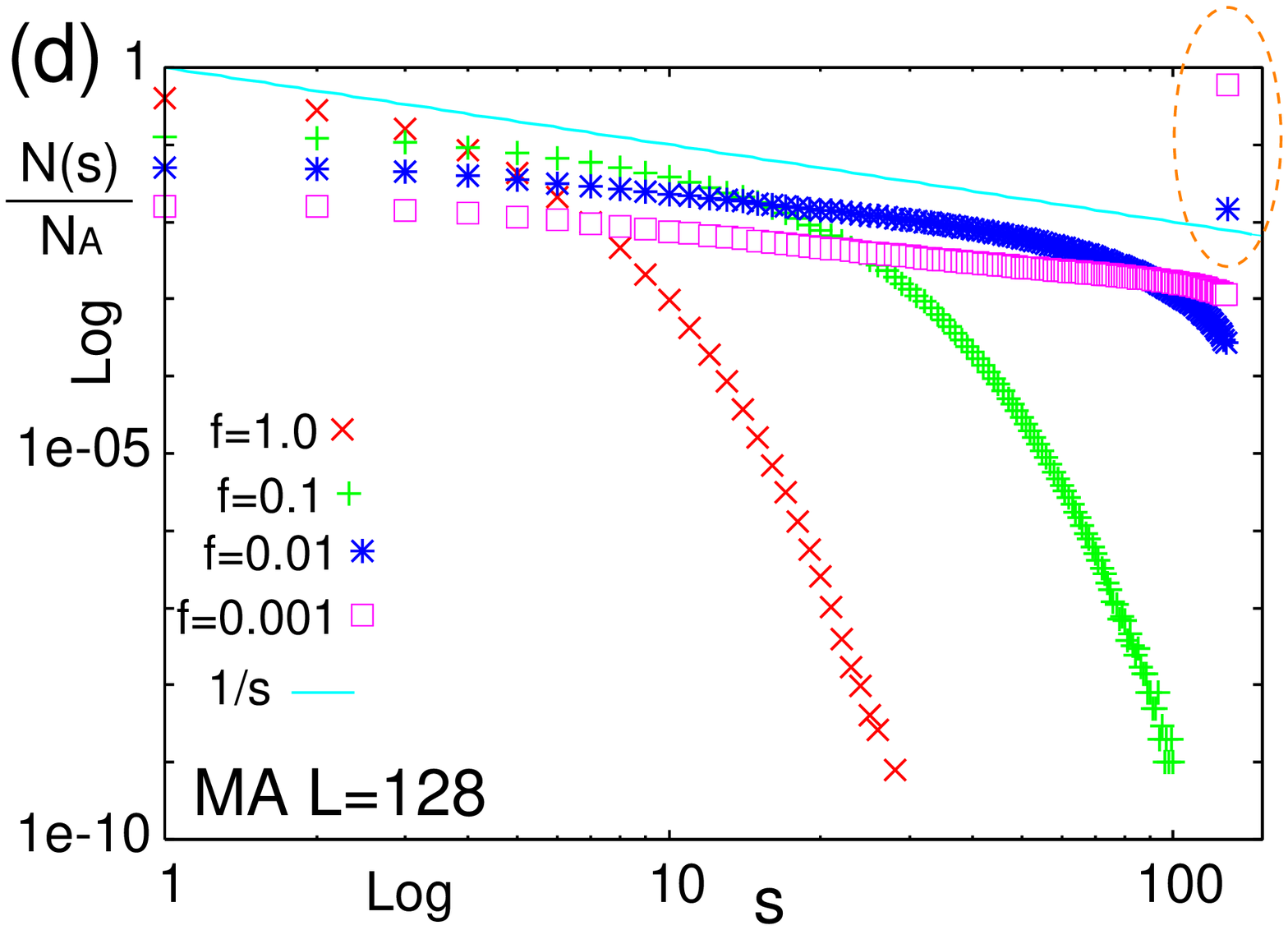}
\caption{Frequency-size distribution of earthquakes for the models M1, M2, M3, and MA, with $L=128$, $p=1.0, f=1.0, 0.1, 0.01$, and $0.001$. The solid line denotes
 $1/s$. The system-size earthquakes are highlighted by the
 orange dotted ellipses. $N_A$ is the total number of earthquakes.}
\end{center}
\end{figure}

Previous studies have found three types of behaviors (``phases'') in the
size-frequency distribution in the spring-block models and cellular
automaton models of earthquakes \cite{Mai,Mor,Tej,Kaw,Vas}.
Following them, we classify the size-frequency distributions of the
simulated earthquakes into three phases, as follows.
When the size-frequency distribution, except for the system-size
earthquakes, can be expressed by a power-law (the GR law), the behavior of the
distribution is called ``critical''.
When the frequencies of large earthquakes, whose sizes are close to $L$,
are higher than would be expected by the power law, it is called ``supercritical''.
When the frequencies of large earthquakes, whose sizes are close to $L$,
are lower than would be expected by the power law, it is called ``subcritical''.
Our definition of phases corresponds to the definitions of other
studies, if we exclude the system-size earthquakes.
These behaviors in the size-frequency relations of earthquakes have been 
observed in different areas and in different faults \cite{Mai}.
These phase descriptions are still debated in the seismological
community.

The phases of the size-frequency distributions thus classified are
summarized in Table I.
For M1, the size-frequency distributions show the critical
phase, independent of $f$; see Fig.~3(a).
The frequency of the system-size earthquakes, which is highlighted by an
orange ellipse in Fig.~3(a), is larger when $f$ is smaller.
This is because, for small $f$, all the sites tend to be loaded before triggering.
The exponent of the decay of frequency with increasing $s$ is
approximately $-1.0$ for $f=0.1$ and $0.01$.
The decay rate is larger when $f$ is larger, because smaller earthquakes
occur more frequently.

In Fig.~3(b), we see that for M2, the supercritical phase is observed for $f=1.0$ and
$0.1$, where the frequency of earthquakes increases with $s$ for
$s\gtrsim 100$, while the critical phase is observed for $f=0.01$ and
$0.001$.
Significant peaks at $s=L$ are observed for all values of $f$.
For M3, we observe peaks at $s=32, 96$, and $128$ for
$f=1.0, 0.1$, and $0.01$; see Fig.~3(c).
These peaks correspond to the distances between the trigger sites.
The peaks at $s=32$ and $96$ tend to be unclear for smaller $f$, because
more sites tend to be loaded before triggering and so the system-size
earthquakes become prominent.
The peaks of frequencies at $s=32$ and $96$ in M3, and the
supercritical behavior for M2 and M3, may be explained as follows:
Small earthquakes occur frequently and release stress near the
trigger sites, and clusters of loaded sites tend to be generated between
the trigger sites.
These clusters correspond to a high frequency of large earthquakes.

For MA, the subcritical phase is observed for all cases; see Fig.~3(d).
The size-frequency distributions may be approximated by power-laws for
small $s$.
With an increase in $s$, the frequency decreases more rapidly than would be expected from
the power law, and this rapid decrease starts at smaller values of $s$ for larger values of
$f$.
The peak in the frequency of earthquakes for $s=L$ was found for $f=0.01$ and
$0.001$, although it was not found for $f=1.0$ and $0.1$.

\begin{table}
\begin{center}
\begin{tabular}{|c|c|c|c|c|}
\hline
 & f=1.0 & f=0.1 & f=0.01 & f=0.001\\
\hline
M1 & $critical$ & $critical$ & $critical$ & $critical$\\
\hline
M2 & $super$ & $super$ & $critical$ & $critical$\\
\hline
M3 & $super$ & $super$ & $super$ & $critical$\\
\hline
MA & $sub$ & $sub$ & $sub$ & $sub$\\
\hline
\end{tabular}
\caption{Phases of the size-frequency distributions in
 Fig.~3, where $super$, $sub$, and $critical$ denote the supercritical,
 subcritical, and critical phases, respectively.}
\label{tab}
\end{center}
\end{table}

\subsection{Scaled LDF for system-size earthquakes}

Next, we calculated the LDFs for the frequencies of the system-size
earthquakes for the models M1, M2, M3, and MA, while varying the system size
$L$ and the triggering rate $f$.

We first evaluated the mean frequency of
the system-size earthquakes $x(L)$ by using the number of system-size
earthquakes that occurred during the simulation; see eq.(\ref{meanxs}).
Figure~4 shows $x(L)$ versus the system size for $f=1.0, 0.1$,
and $0.01$, and for models M1, M2, M3, and MA.
For M3, the trigger sites are located at both the ends and at $m=L/2+1$
for even values of $L$ and $m=(L+1)/2$ for odd values.
Note that the results for M1, M2, and M3 are plotted on logarithmic
coordinates in Fig.~4(a), while those for MA are on semilogarithmic
coordinates in Fig.~4(b).
The time step was taken as $2^{24}$ in each case.
For M1, M2, and M3, with $f=1.0$ and $0.1$, $x(L)$ decreased
as $L$ increased.
When $f=0.01$, $x(L)$ seems to be independent of $L$, as shown
in Fig.~4(a), although for M1, it approximately obeys power-law decay with an
exponent of about $-0.01$.
For MA and $f=1.0$, $x(L)$ exponentially decreases with
increasing $L$.
For $f=0.1$ and $0.01$, $x(L)$ takes the maximum values at $L\sim 5$ and
$L\sim 27$, respectively, and exponentially decreases with $L$ for large
$L$.
The decrease of $x(L)$ with an increase in $L$ occurs because the
preparation time for a system-size earthquake increases with $L$.
\begin{figure}
\begin{center}
\includegraphics[scale=0.3]{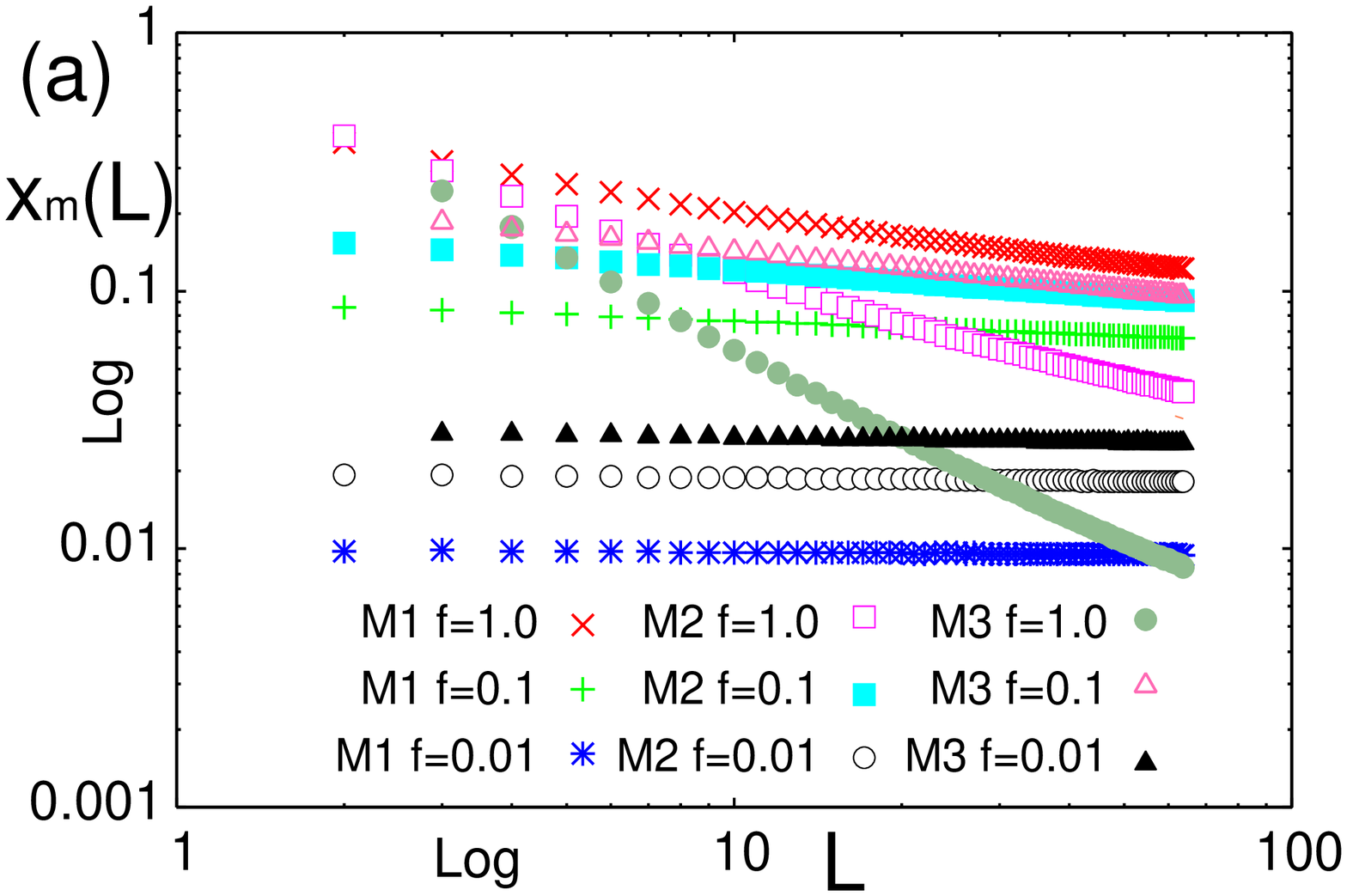}
\includegraphics[scale=0.3]{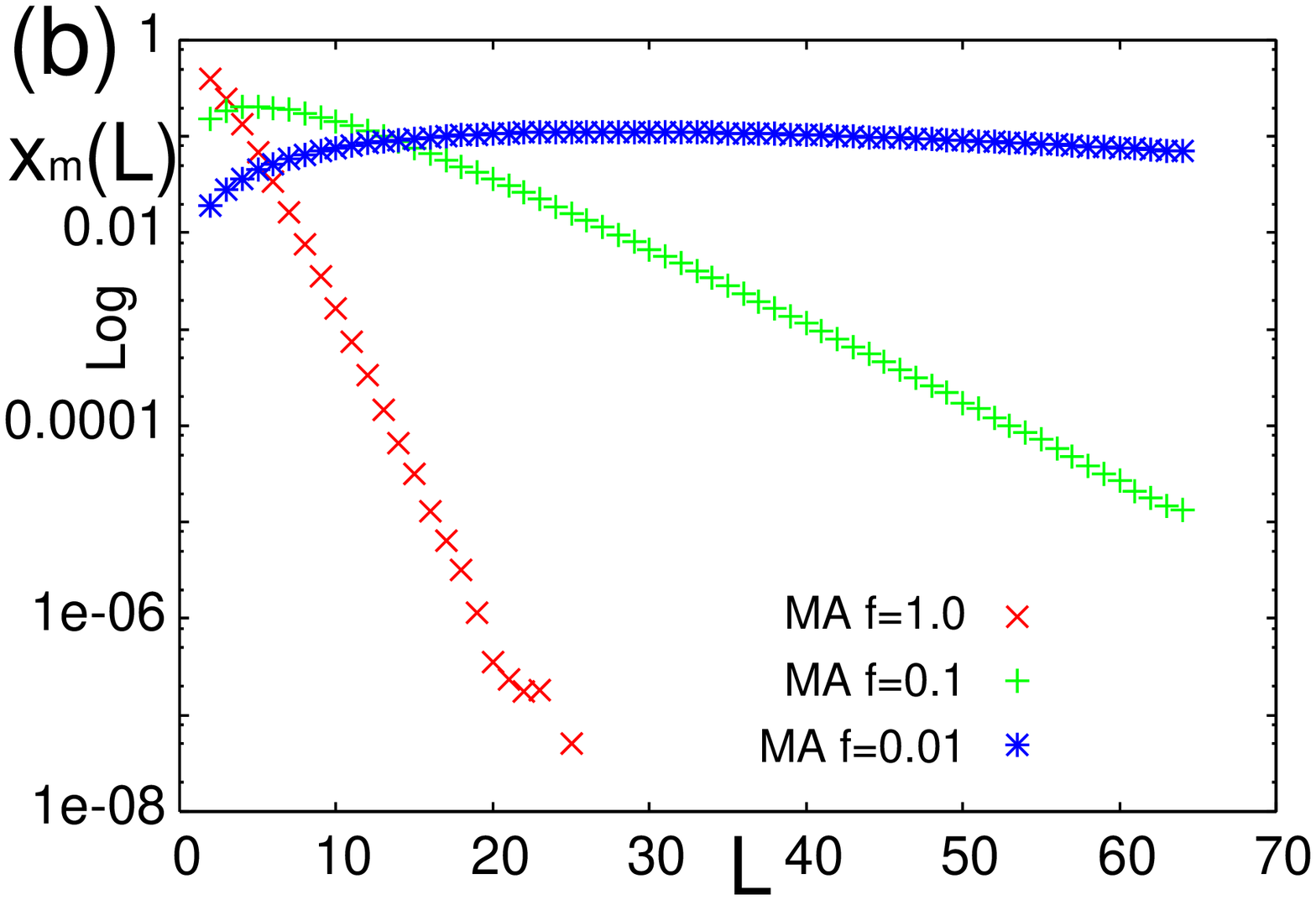}
\caption{Mean frequency of system-size earthquakes $x(L)$ versus
 the system size $L$.
 (a) Models M1, M2, and M3, with $f=1.0, 0.1$, and $0.01$. (b) Model MA
 with $f=1.0, 0.1$, and $0.01$.}
\end{center}
\end{figure}

The mean frequency of the system-size earthquakes can be also estimated
by the LDF, because the LDF has a minimum at $x_m(L)$.
Note that $x_m(L)$ can be well approximated by $x(L)$ for large $t$.
To calculate $x_m(L)$ for each case, we
 numerically calculated the LDF and fit it with a polynomial, using the
 least-squares method. We then used the $x$ that minimizes this polynomial
 as $x_m(L)$.
The order of the polynomial was taken to be $20$.
The $x(L)$ that was calculated with a single simulation run was obtained much more quickly
but was less accurate 
than the $x_m(L)$ calculated from the LDF.
For the parameters used to calculate the LDFs in this study, $|x(L)-x_m(L)|/x(L)$ is smaller than $3.2\%$.

Because $x_m(L)$ depends on $L$, appropriate
scaling is required to compare the LDFs for systems that have different sizes.
We used $x_m(L)$ to scale the LDF empirically.
To introduce scaling by $x_m(L)$, which is simply written as $x_m$
hereinafter, we define a scaled variable $z$ as
\begin{equation}
 z=\frac{x-x_m}{x_m}.
\end{equation}
In this scaling, the LDF of the Poisson process is given by 
\begin{equation}
\label{ldfpoissonsc}
 \frac{\phi_P(z)}{x_m}=(z+1)\log{(z+1)}-z.
\end{equation}
The division by $x_m$ on the left-hand side of (\ref{ldfpoissonsc}) is
 unnecessary if the time $t$ is scaled by $x_m$, and the LDF written by
 $z$ omits $x_m$ and can simply be written as $\phi_P(z)$. 
The introduction of $z$ enables us to compare more clearly the calculated LDF to the LDF
of the Poisson process (\ref{ldfpoissonsc}).

The scaled LDFs for the numerically calculated frequencies of
system-size earthquakes are shown in Fig.~5.
The system size $L$ varies as $L=8, 10, 12, 16$, and $32$.
The LDFs in the cases of $L=16$ and $32$ are obtained by the cloning Monte
Carlo method, while the others are by the matrix method.
It is difficult to calculate the LDF for MA with $f=1.0$ and $L=16$ and
$32$ because the number of system-size earthquakes is very small, and so
instead, the LDF for $L=13$ is plotted.
The red solid curve in each panel denotes the LDF of the Poisson process
$\phi_P(z)$ (\ref{ldfpoissonsc}), the blue solid curve  denotes the scaled
LDF of the one-site system $\phi_1(z)$ (\ref{onesite}), and the black
dashed curves in (a) and (e) denote
reference quadratic functions whose coefficients were determined by
eye.
In calculating $\phi_1(x)$, we introduced the effective triggering rate
$f'=n_tf$ of the system-size earthquake, 
where $n_t$ is the number of trigger sites $n_t$.
When the LDF is a quadratic function, $P(x)$ is Gaussian and the
fluctuation is characterized only by the variance.
Scaling by $x_m$ worked well for some cases.
The scaled LDFs for different values of $L$ collapse onto a
curve for M1 with $f=0.01$ (Fig.~5(c)) and for MA with $f=1.0$
(Fig.~5(j)).
In contrast, in the other cases, the scaled LDFs for different values of $L$
are scattered.

\begin{figure}
\begin{center}
\includegraphics[scale=0.25]{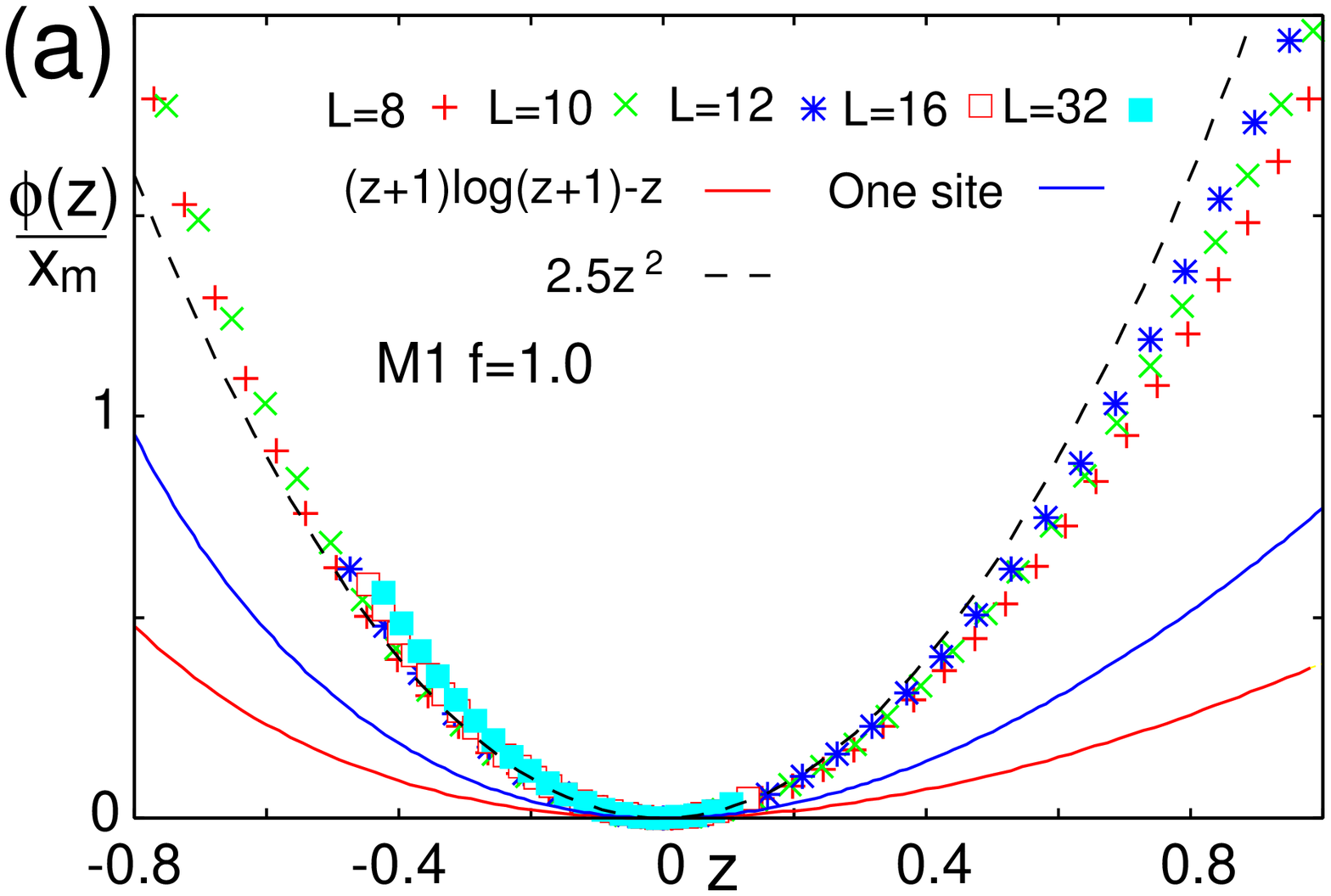}
\includegraphics[scale=0.25]{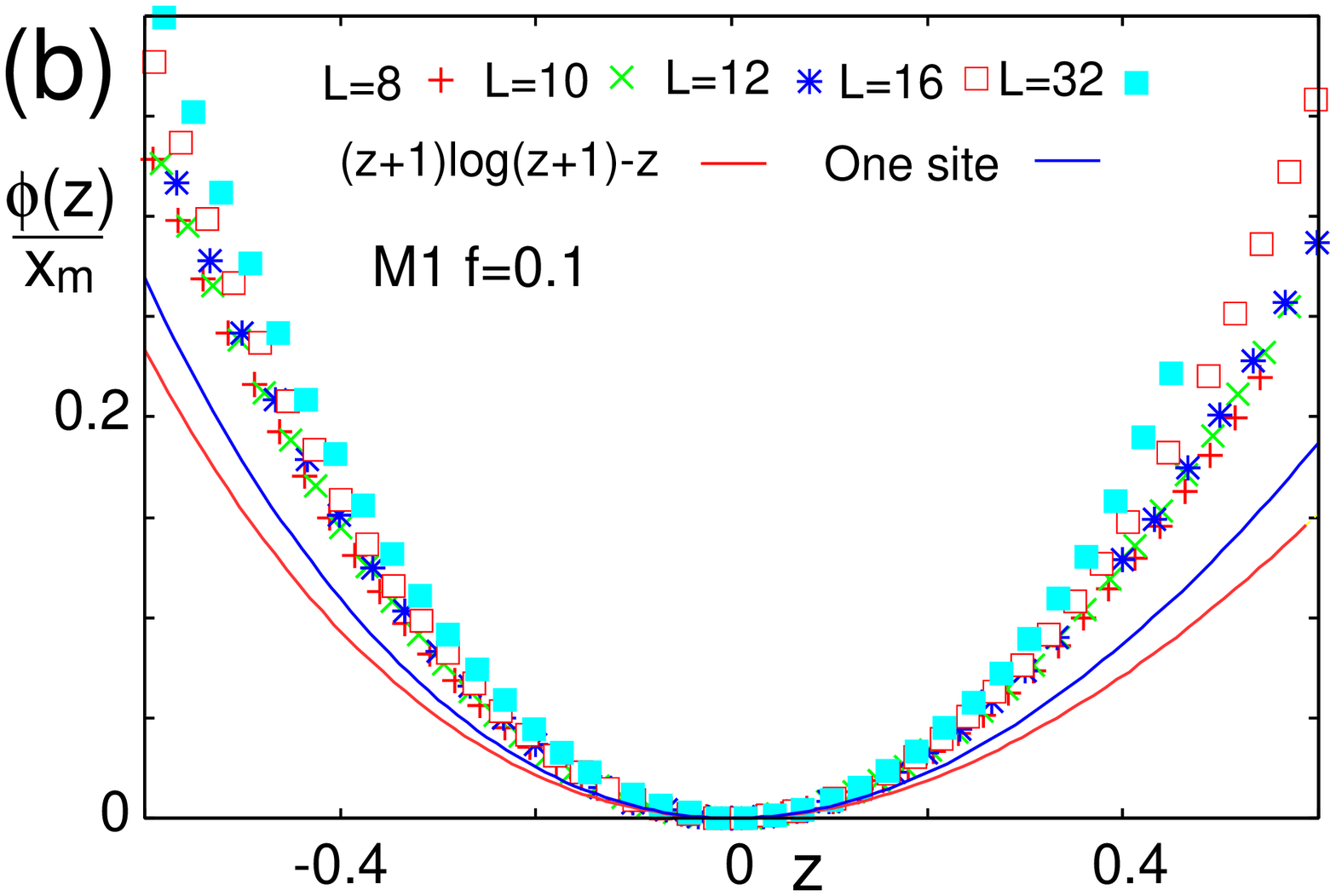}
\includegraphics[scale=0.25]{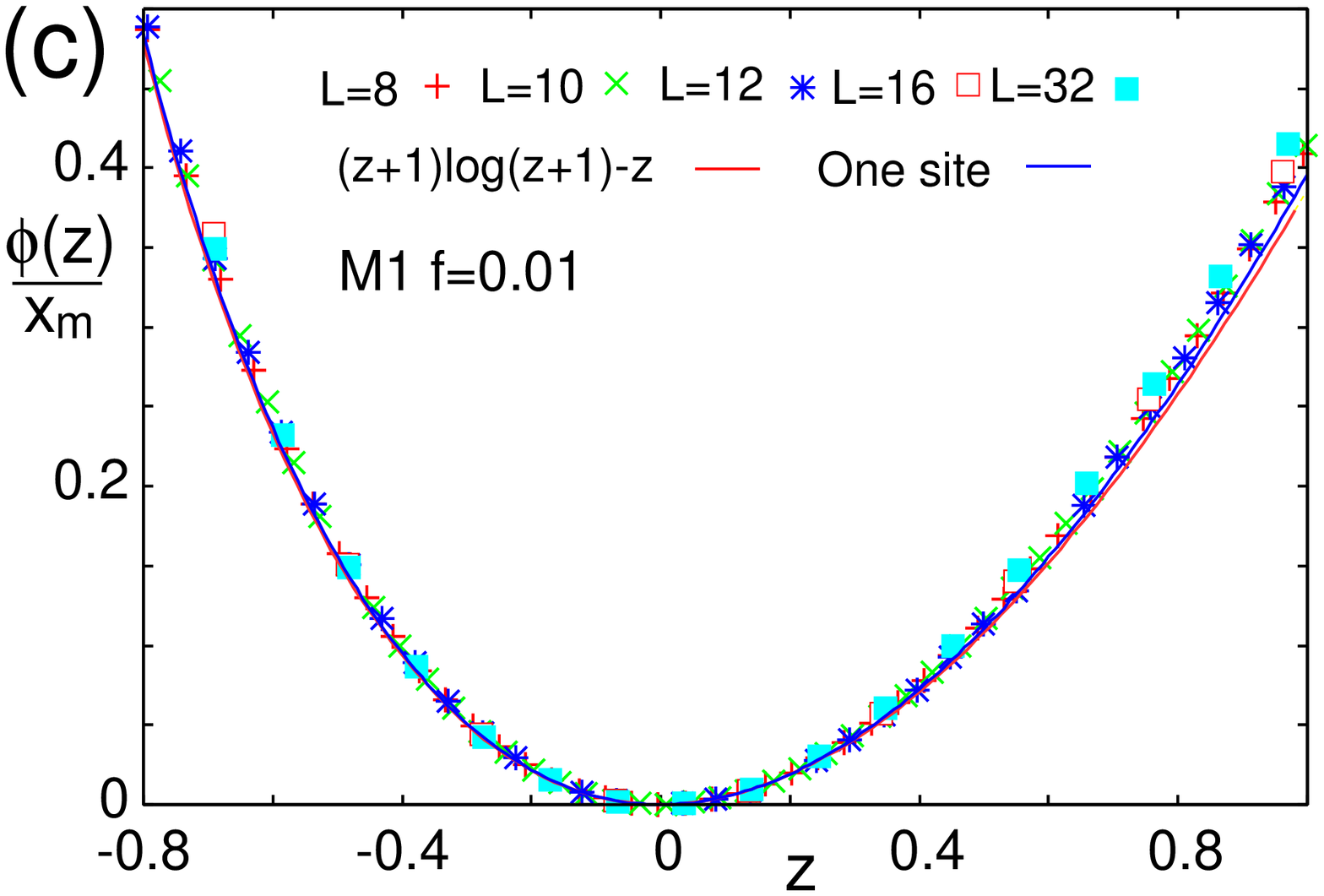}
\includegraphics[scale=0.25]{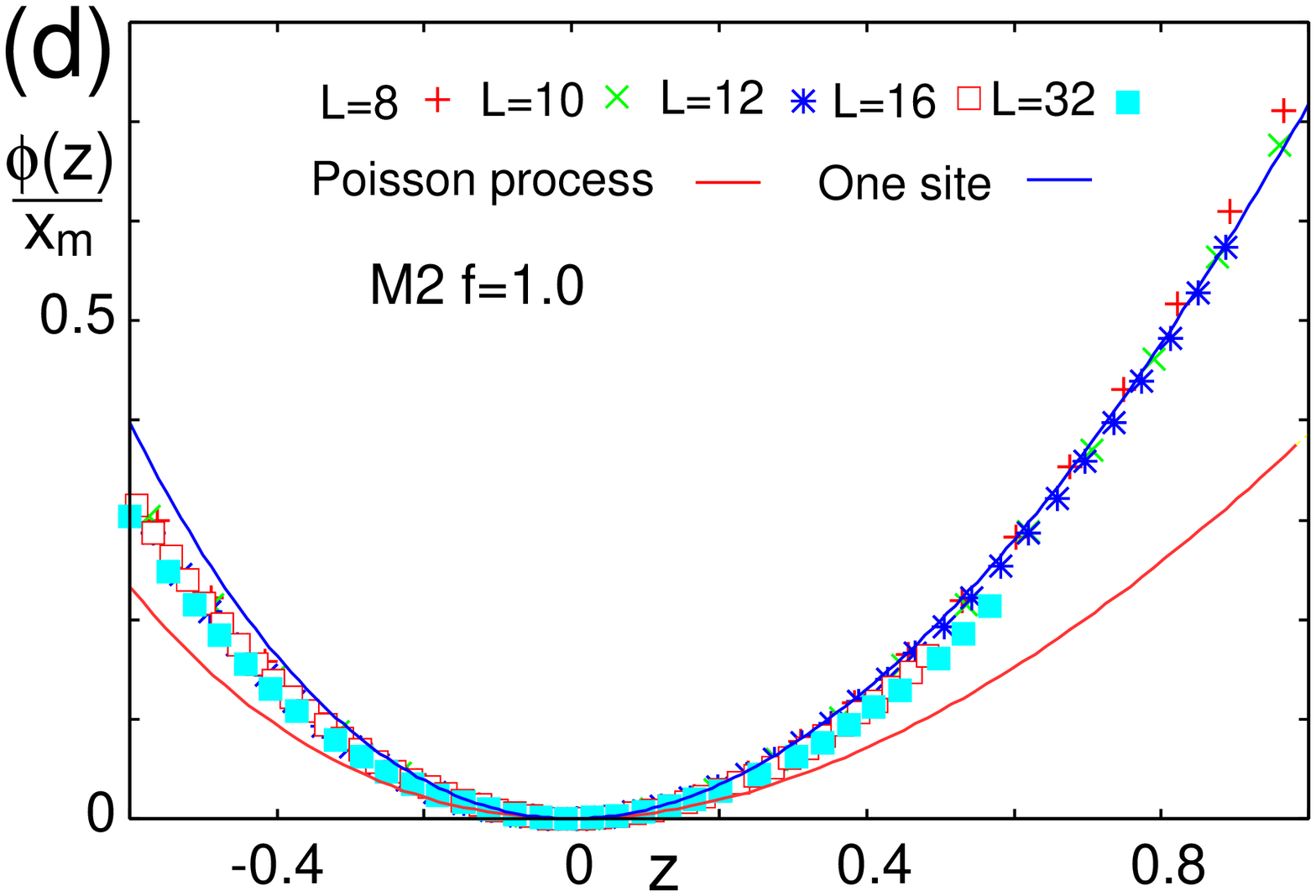}
\includegraphics[scale=0.25]{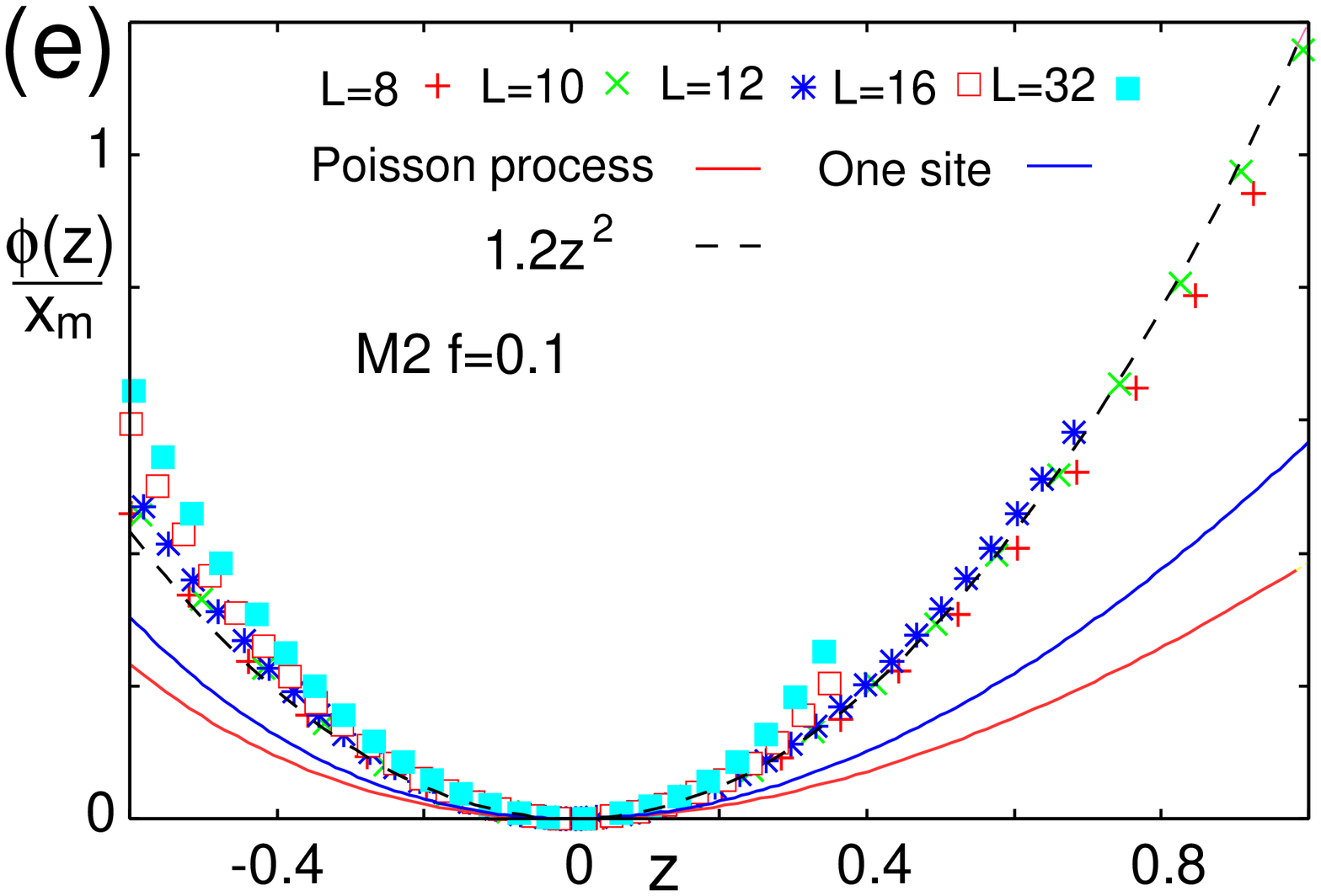}
\includegraphics[scale=0.25]{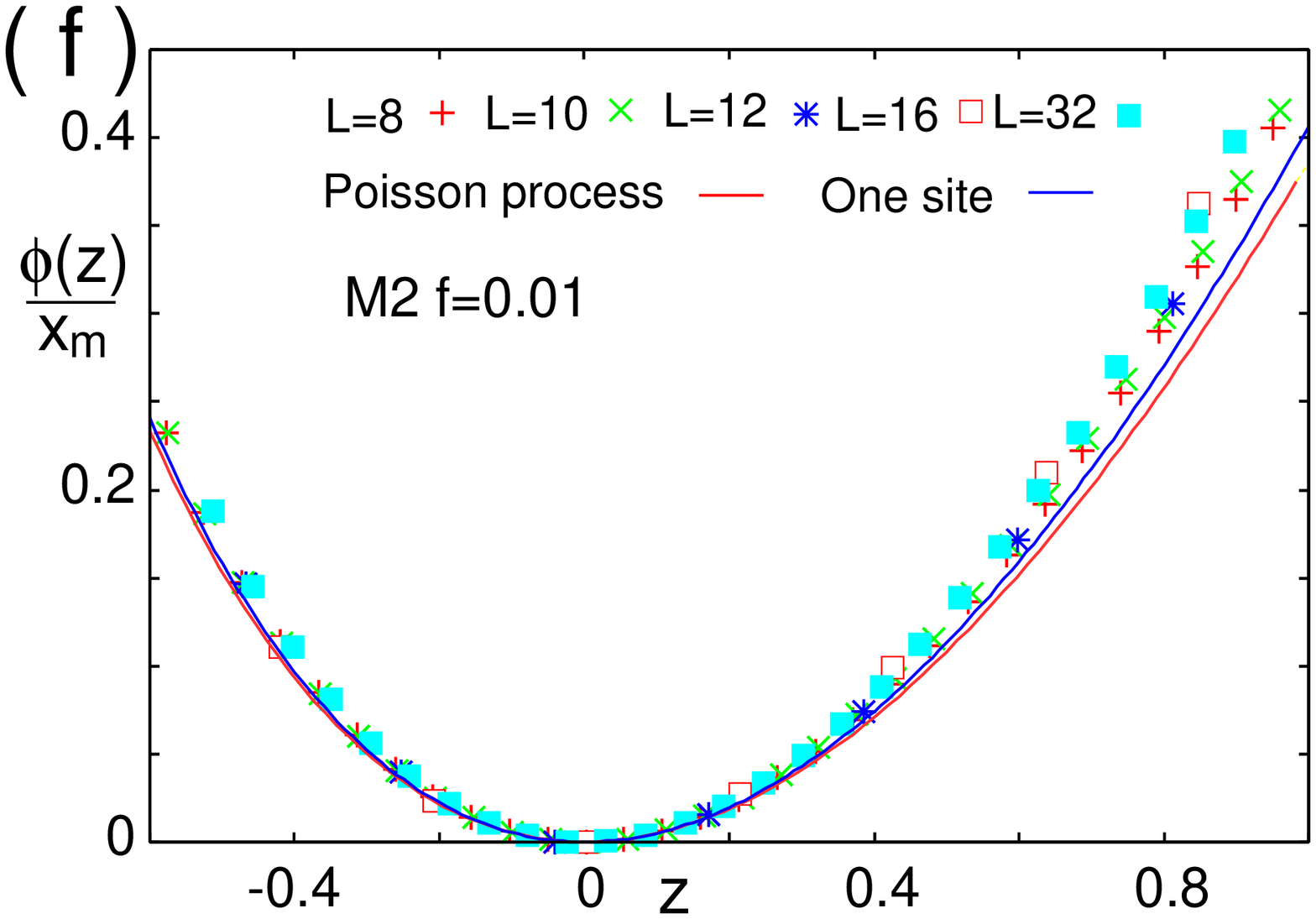}
\includegraphics[scale=0.25]{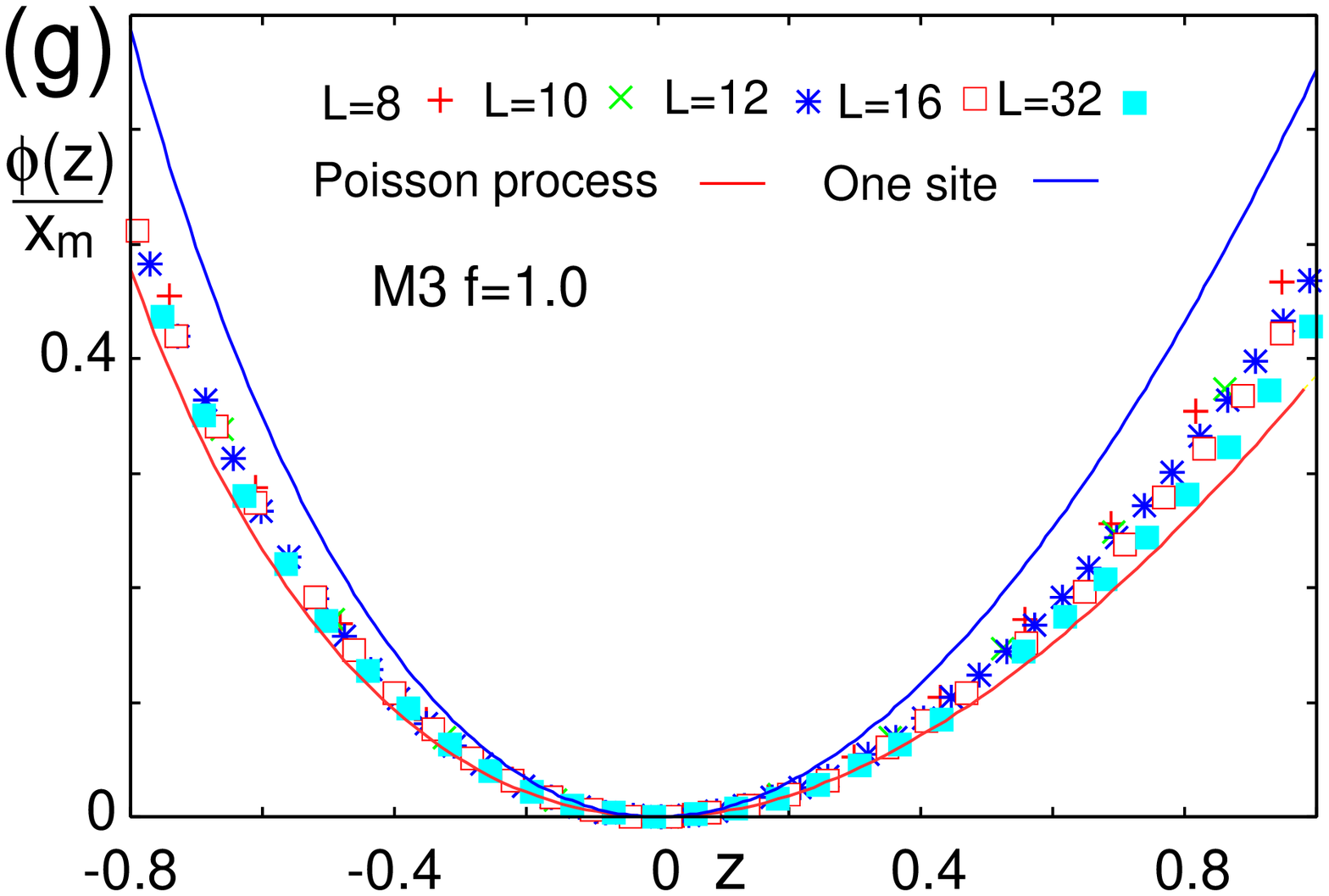}
\includegraphics[scale=0.25]{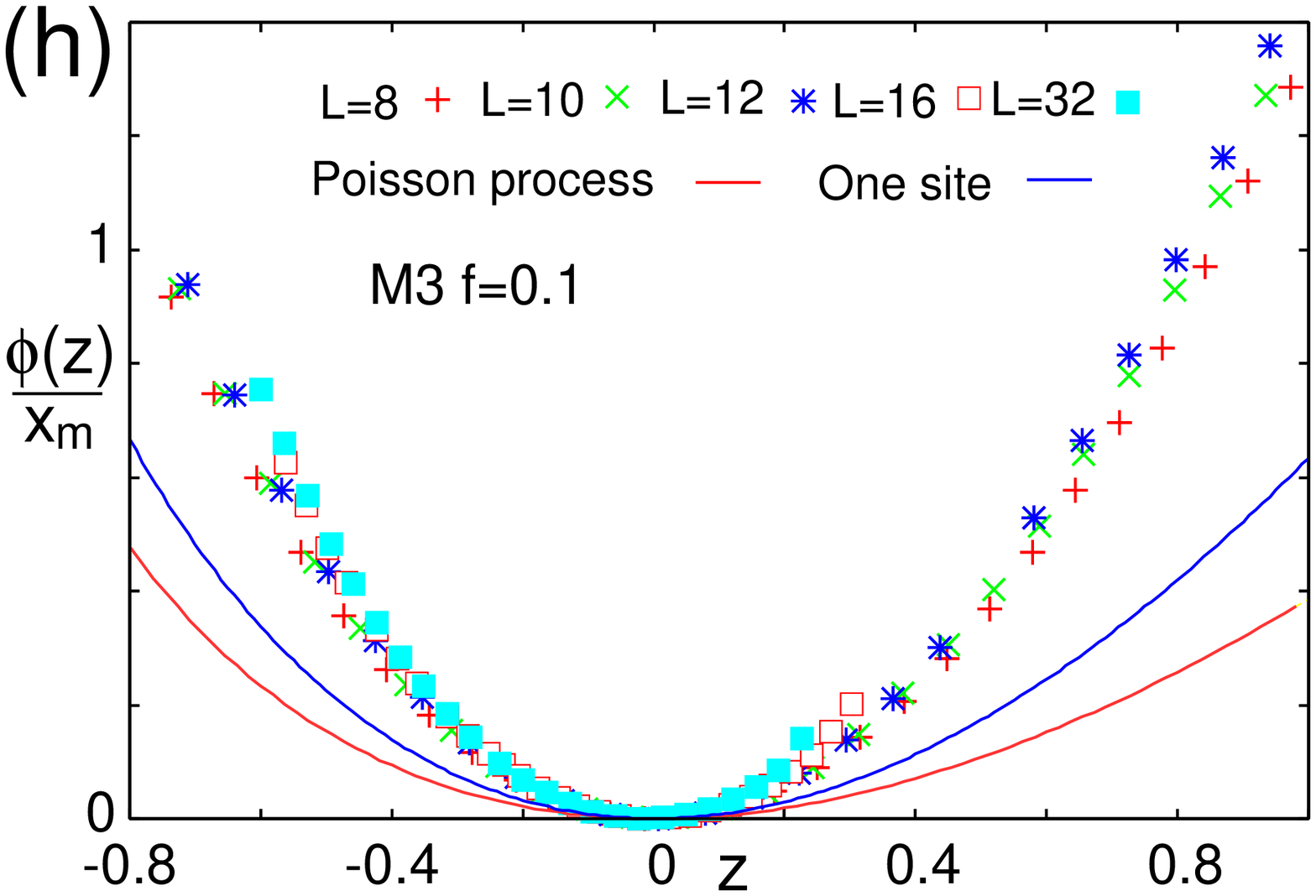}
\includegraphics[scale=0.25]{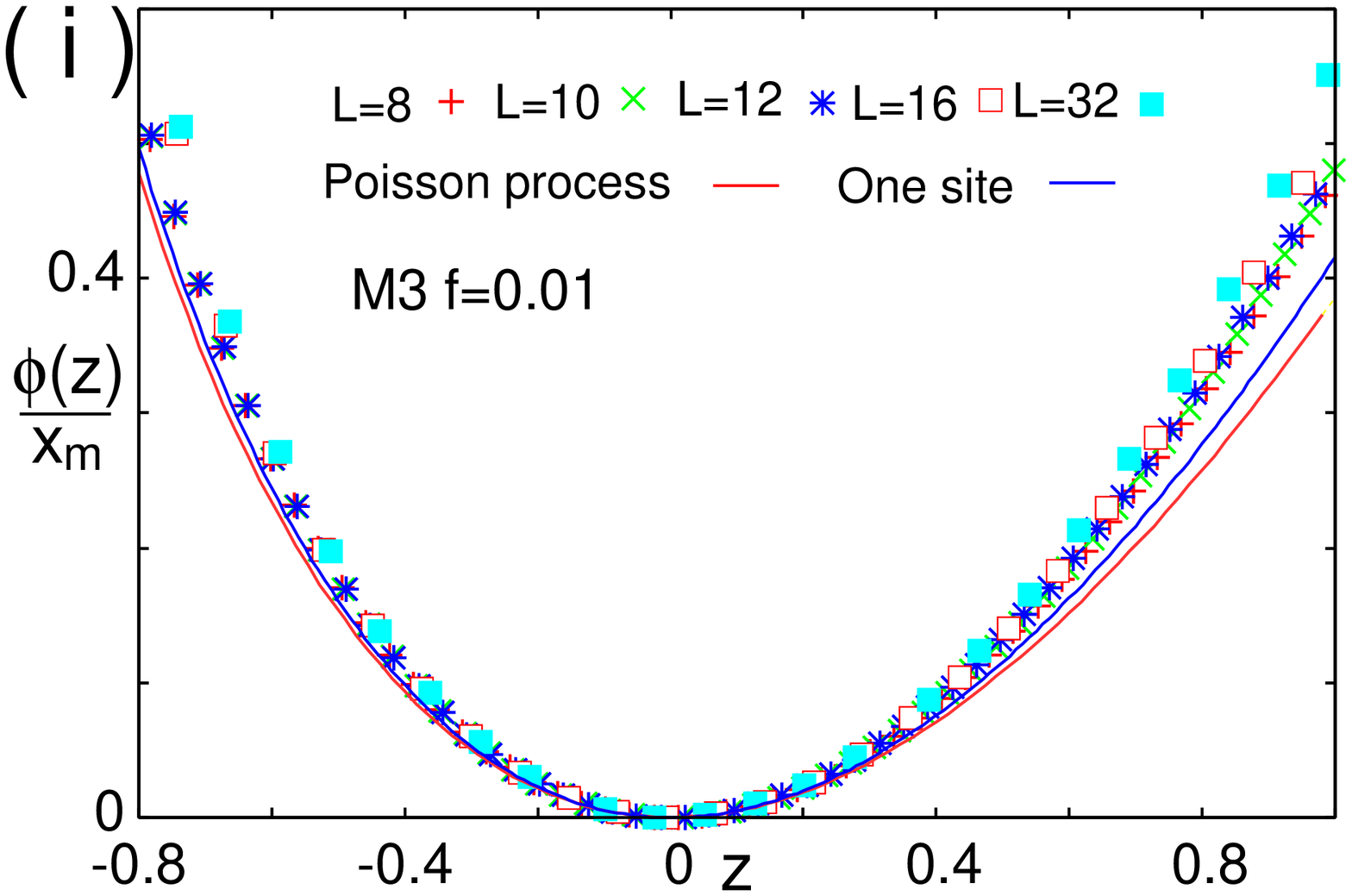}
\includegraphics[scale=0.25]{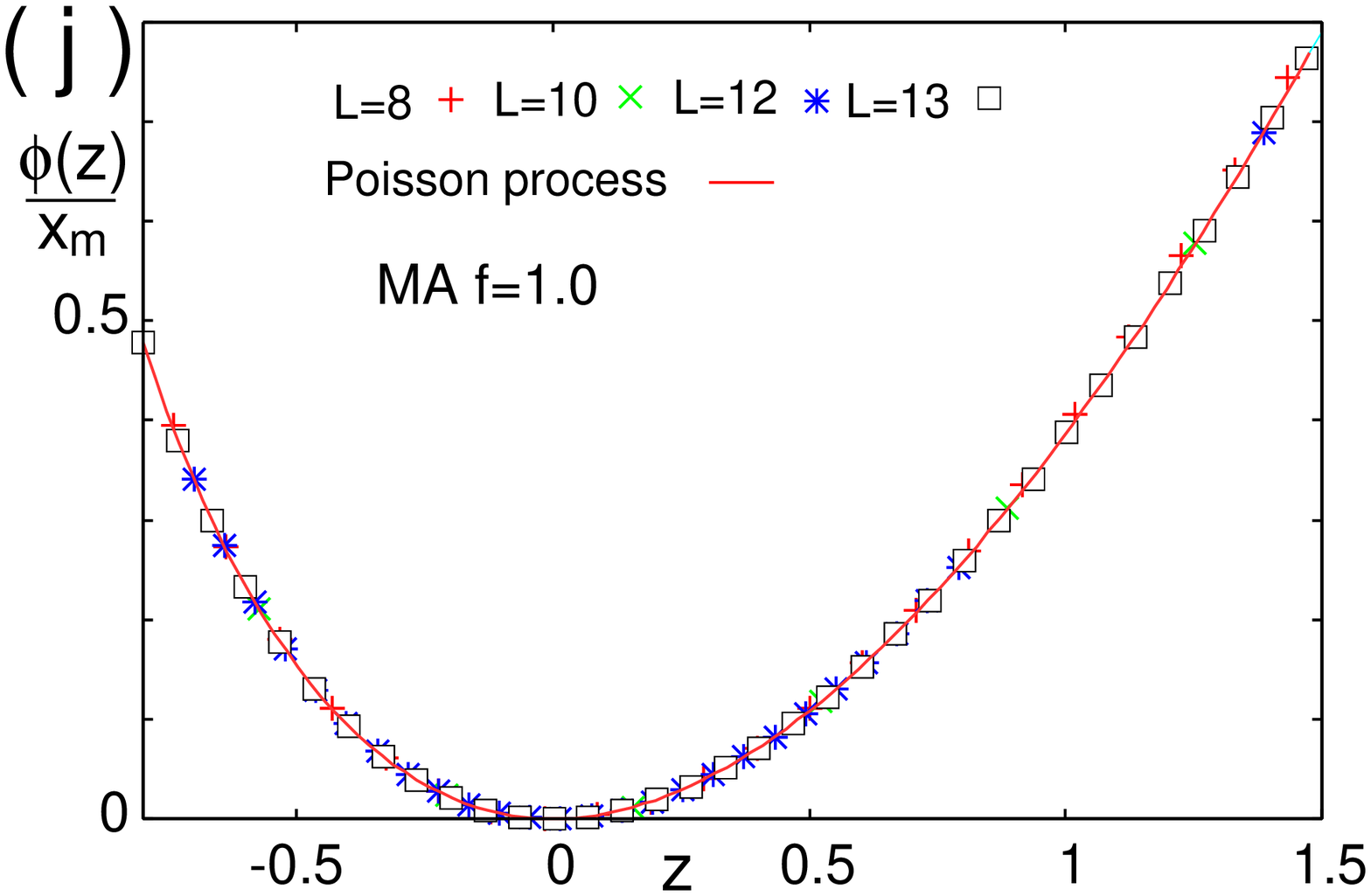}
\includegraphics[scale=0.25]{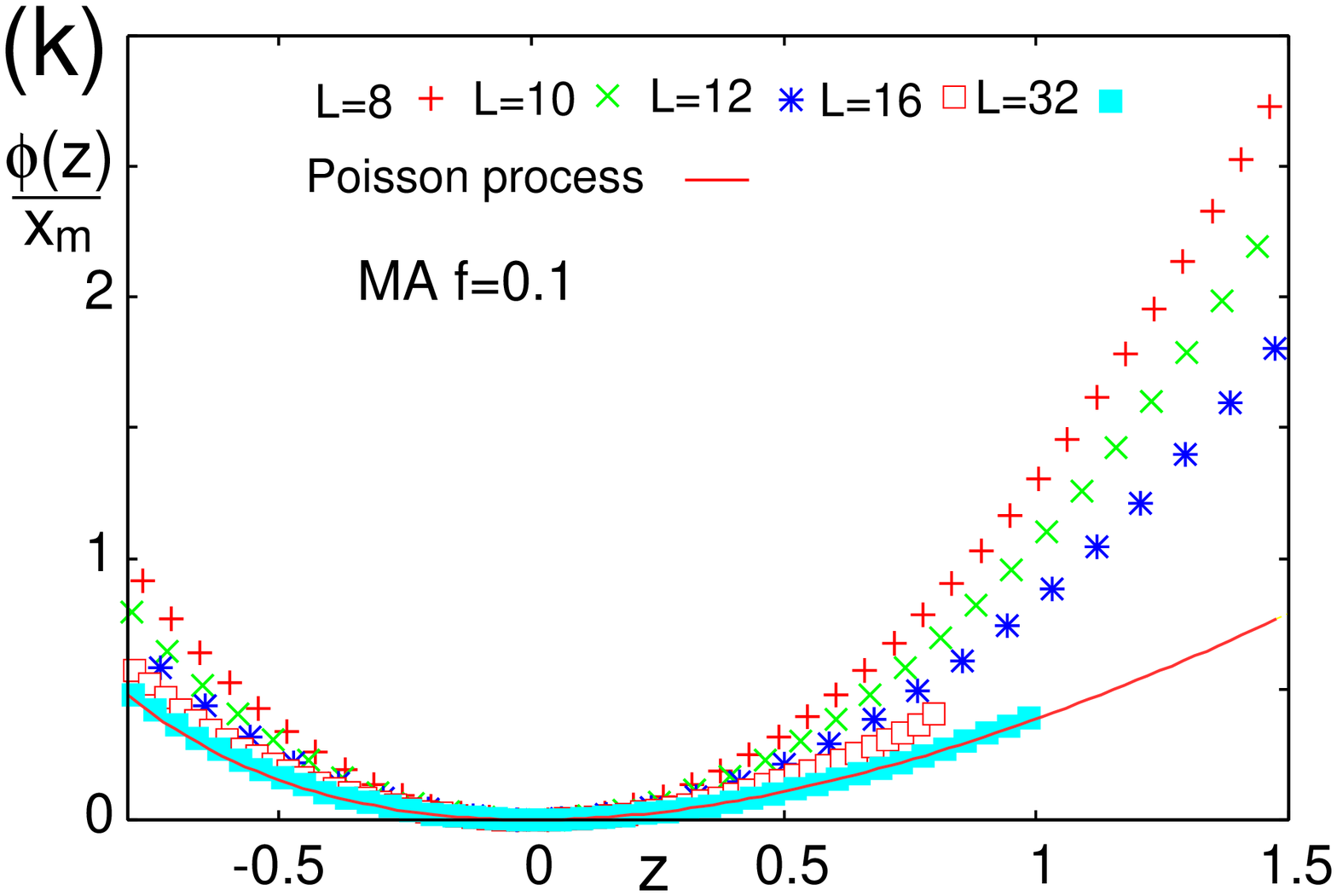}
\includegraphics[scale=0.25]{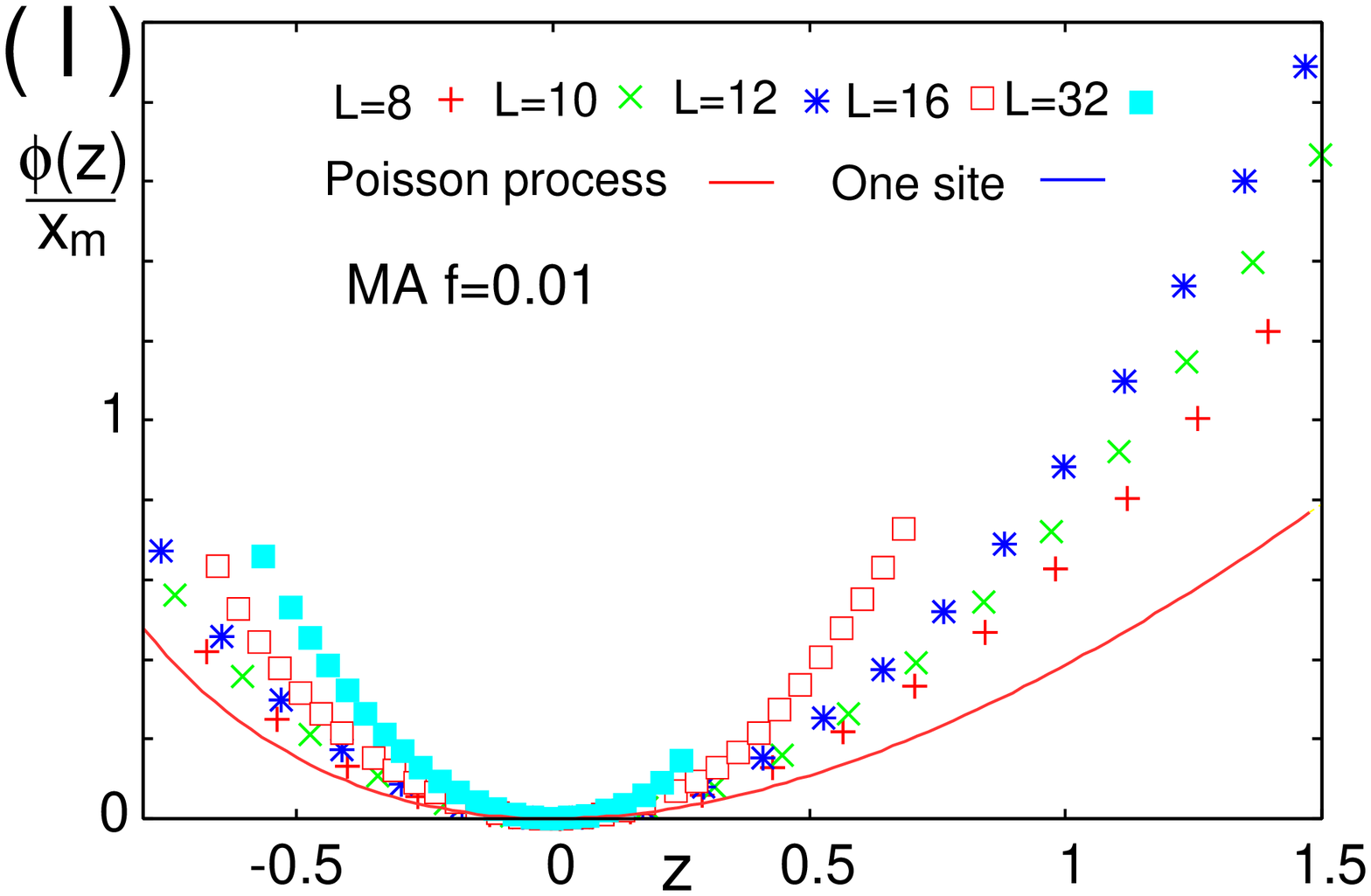}
\caption{The plot of $\phi(z)$ estimated by two methods for models
 M1 ($f=1.0$(a), $0.1$(b), $0.01$(c)), M2 ($f=1.0$(d), $0.1$(e), $0.01$(f)),
 M3 ($f=1.0$(g), $0.1$(h), $0.01$(i)), and
 MA ($f=1.0$(j), $0.1$(k), $0.01$(l)).}
\end{center}
\end{figure}

From a comparison of the LDFs for the system-size earthquakes here obtained
with those of the Poisson process, we find three types of behaviors:
Poisson, two-state, and non-Poisson.
The LDFs for MA with $f=1.0$ are well approximated by the LDF
of the Poisson process, independent of $L$ (Fig.~5(j)); this is called
the ``Poisson phase''.
In this case, the occupation of all sites rarely occurs because of
frequent triggering.
For M1, M2, and M3 with $f=0.01$, $\phi_P(z)$
reasonably approximates the simulated LDFs for $z<0$, while it deviates from the
LDFs for $z>0$ (Figs.~5(c), (f) and (i)).
We call this behavior the ``two-state phase''.
In the two-state phase, the plots are
better approximated by $\phi_1(z)$ than by $\phi_P(z)$.
$\phi_1(z)$ is close to $\phi_P(z)$ for $z<0$, though it
deviates from $\phi_P(z)$ for $z>0$.
This is because earthquakes of for which $s<L$ rarely occur for
small $f$, and the system-size earthquakes are dominant.
When the frequency of system-size earthquakes is much larger than that
of the other earthquakes, the system is approximated by the transition between 
two states $\{0,\cdots,0\}$ and $\{1,\cdots,1\}$, which is similar to the
one-site forest-fire model of two states $\{0\}$ and $\{1\}$.
For $z<0$, the number of system-size earthquakes during a time interval
$t$ is small compared to the mean, and the time intervals between successive
system-size earthquakes are longer than the average.
When the time interval is long, all the sites tend to be loaded before
triggering.
This results in a nearly random occurrence of system-size earthquakes,
 and the LDF may be approximated by the LDF of the Poisson process.
For a homogeneous Poisson process, the occurrence rate of events is constant
with time. The events are distributed randomly along the time axis,
and there is no correlation between events.
We note that the origin of the Poisson behavior in the two-state
phase for $z<0$ is different from that of the Poisson phase, in that
in the two-state phase, it originates in the
Poisson process of triggering, while in the Poisson phase, it
originates in the occurrence of full loading of the system.

For the other cases, the simulated LDFs clearly deviate from that of the
Poisson process, and this is called the ``non-Poisson phase''.
Although the LDFs show non-Poisson behavior for MA with $f=0.1$ and
$0.01$ (Figs.~5(k) and (l)), we anticipate that the distributions converge
to the Poisson curve with increasing $L$, as discussed in the next section.
For this reason, we write N(P) for MA with $f=0.1$ and $0.01$.
The phases of the LDFs thus classified are shown in Table II.
For M1, M2, and M3, we do not have a clear explanation for the non-Poisson
behavior.
The troughs in the LDFs for M1 with $f=1.0$ and $0.1$ (Figs.~5(a) and (b)),
and for M2 and M3 with $f=0.1$ (Figs.~5(e) and (h)), are deeper for larger
$L$.
In contrast, the troughs in the LDFs for M2 and M3 with $f=1.0$
(Figs.~5(d) and (g)) are shallower for larger $L$.
A deep trough in the LDF indicates that the frequency of system-size
earthquakes is commonly close to $x_m$, and as the trough becomes deeper,
the system-size earthquakes occur more periodically.
A renewal process is characterized by a distribution of time intervals
between successive events (here the events are the system-size
earthquakes), and in a renewal process, the relation between the deep trough in
the LDF and the periodicity are explicitly related to each other. The sequence of the system-size earthquakes in the
present forest-fire models is a renewal process because the system
always becomes empty after each system-size earthquake.
The probability that $N$ system-size earthquakes occur during time $t$,
which is related to the LDF for the frequency of system-size earthquakes, 
depends on the distribution of the time intervals at which the sequence of system-size earthquakes are renewed.
Let us assume an LDF that has a very deep trough 
at $x_m$, and a distribution of time intervals between successive system-size
earthquakes in which the dominant peak is at a time interval $\Delta$.
In this case, $P_{cl}(x)(\sim e^{-t\phi_{cl}(x)})$ is close
to a delta function with the peak at $x_m$, which means that the
probability of the trajectories of system-size earthquakes deviating
from $x=x_m$ is very small.
The trajectories that give $x=x_m$ consist of almost equal time
intervals $\sim\Delta$ or of time intervals that contain deviations from
$\Delta$.
The latter case is unlikely to occur because a deviation from $\Delta$
must be supplemented by other
earthquakes that also deviate from $\Delta$.
\begin{table}
\begin{center}
\begin{tabular}{|c|c|c|c|}
\hline
 & f=1.0 & f=0.1 & f=0.01 \\
\hline
M1 & N & N & T \\
\hline
M2 & N & N & T \\
\hline
M3 & N & N & T \\
\hline
MA & P & N(P) & N(P) \\
\hline
\end{tabular}
\caption{Classification of the LDF by $f$ of the four models. P, N, and
 T denote the Poisson, the non-Poisson, and the two-state phases. The N(P) phase 
 becomes the Poisson phase when $L$ is sufficiently large.}
\label{tab1}
\end{center}
\end{table}

\subsection{Time-interval distributions}

The LDF for the frequency of the system-size earthquakes is related to the
time-interval distribution.
Poisson and non-Poisson behaviors are both observed for the LDF. 
Bunde et.al. studied non-Poisson behavior of the time-interval
distributions by generating long-term correlated data \cite{Bun,Alt}, and
by analyzing several climate records \cite{Bun2}.
They related non-Poisson behavior of the time-interval distribution
of rare events to a stretched exponential function originating in a long-term correlation, and they related Poisson behavior to an exponential decay.
To clarify the cause of the non-Poisson behavior of the LDF,
we calculated the time-interval distribution of the system-size
earthquakes.

In Fig.~6, the time-interval distributions of the system-size
earthquakes, denoted by $\omega(T)$, are presented for
each of the models with $L=8,16$, and $32$, and $f=1.0,0.1$, and $0.01$.
\begin{figure}
\begin{center}
\includegraphics[scale=0.25]{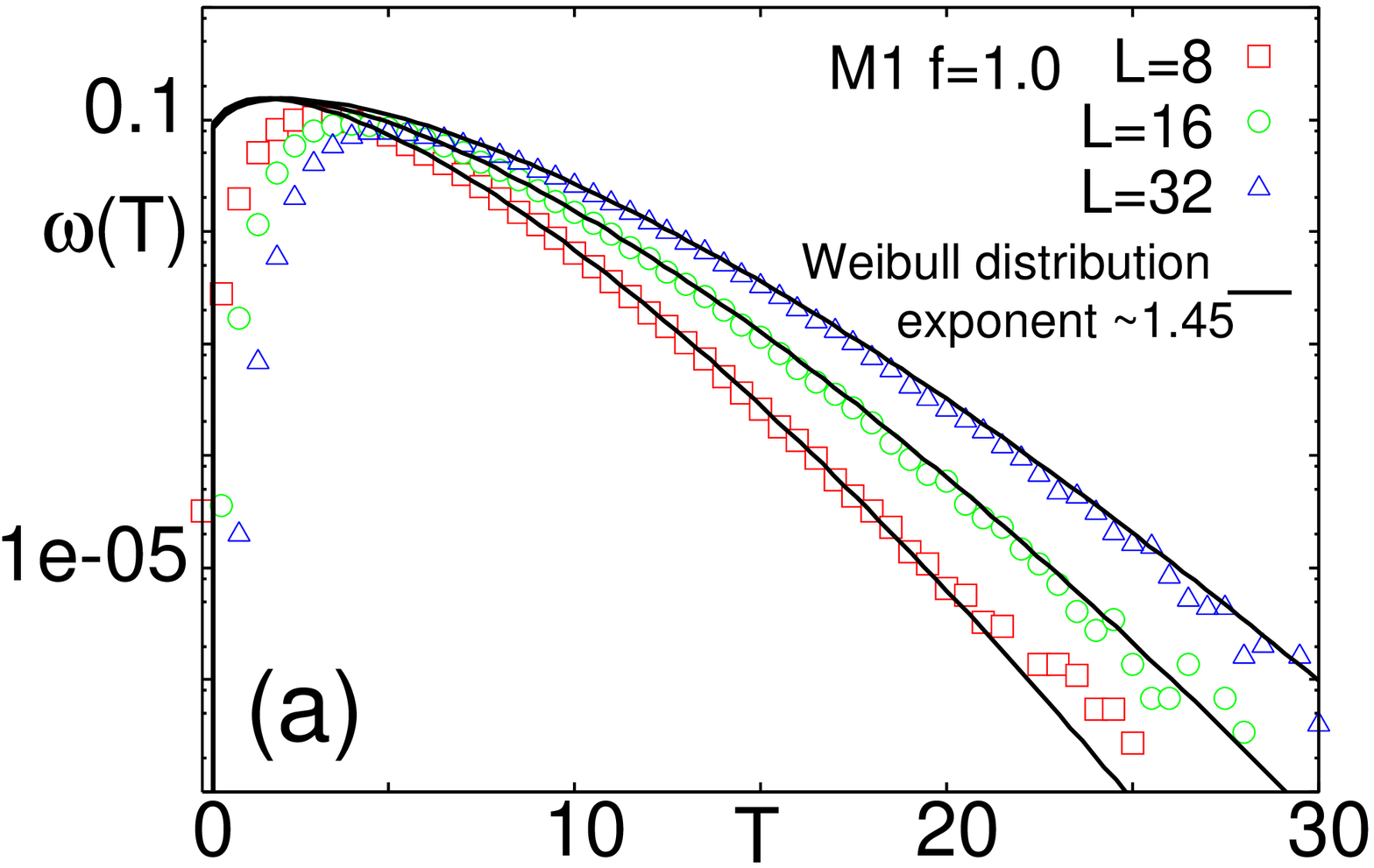}
\includegraphics[scale=0.25]{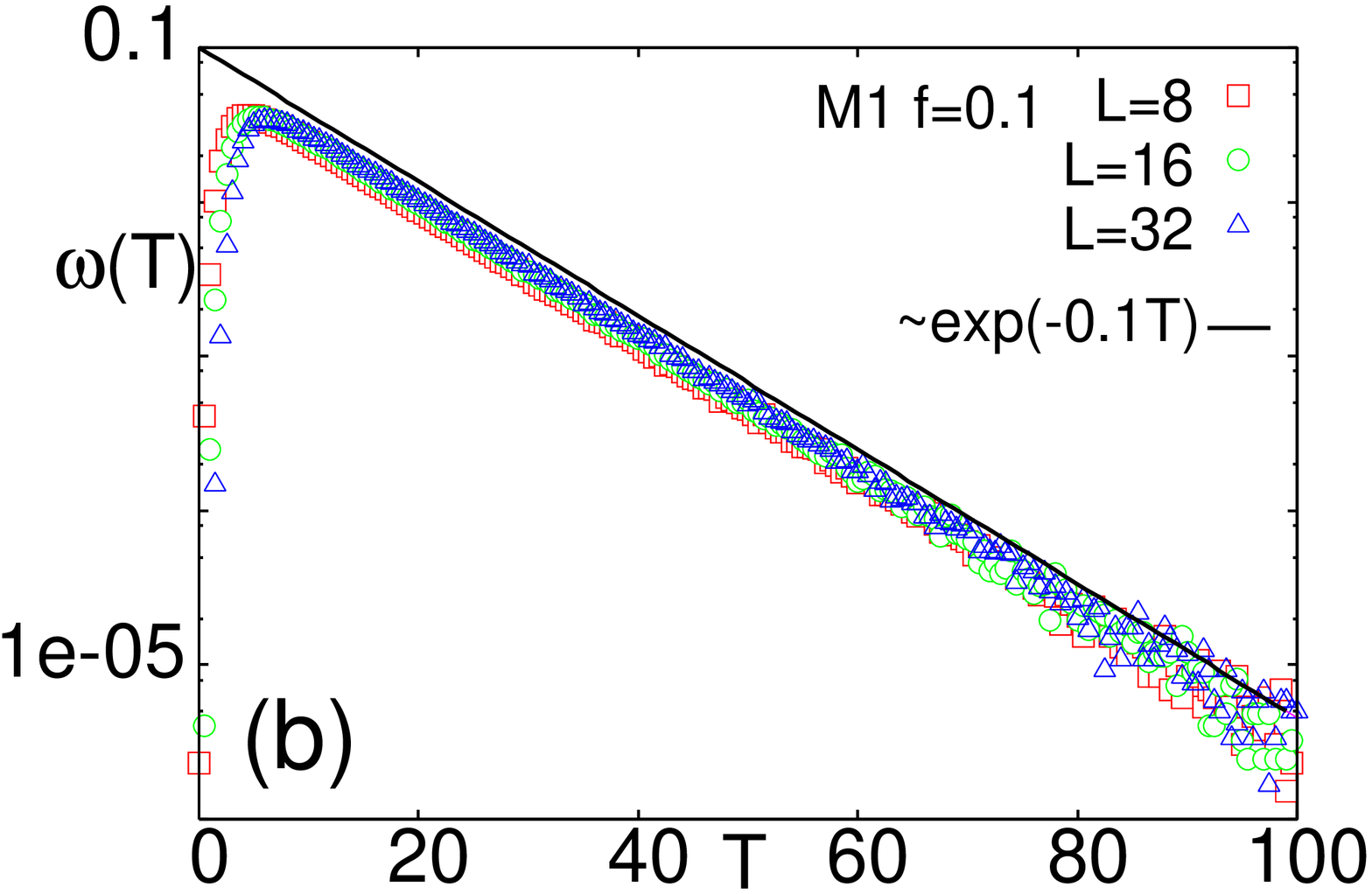}
\includegraphics[scale=0.25]{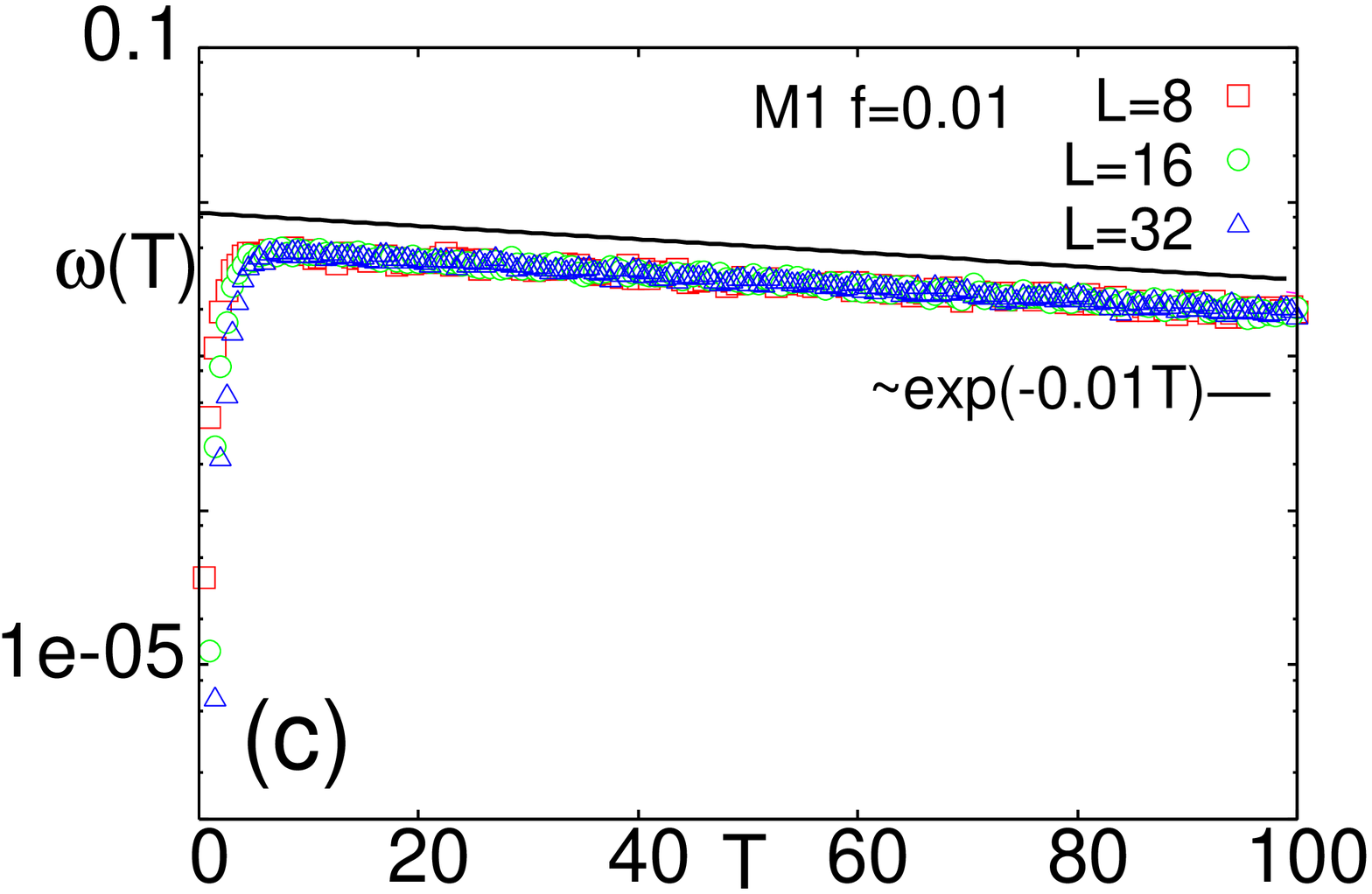}
\includegraphics[scale=0.25]{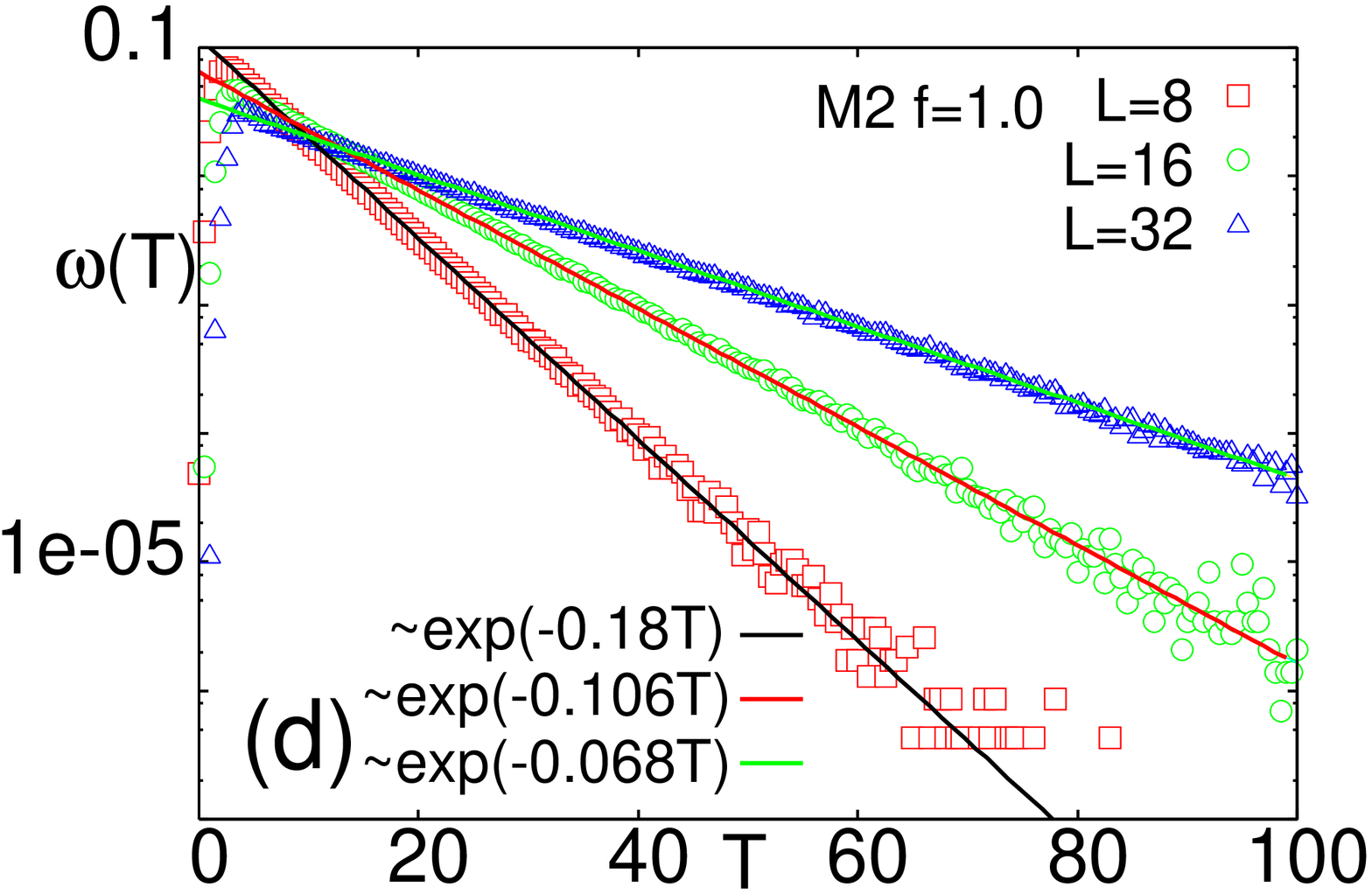}
\includegraphics[scale=0.25]{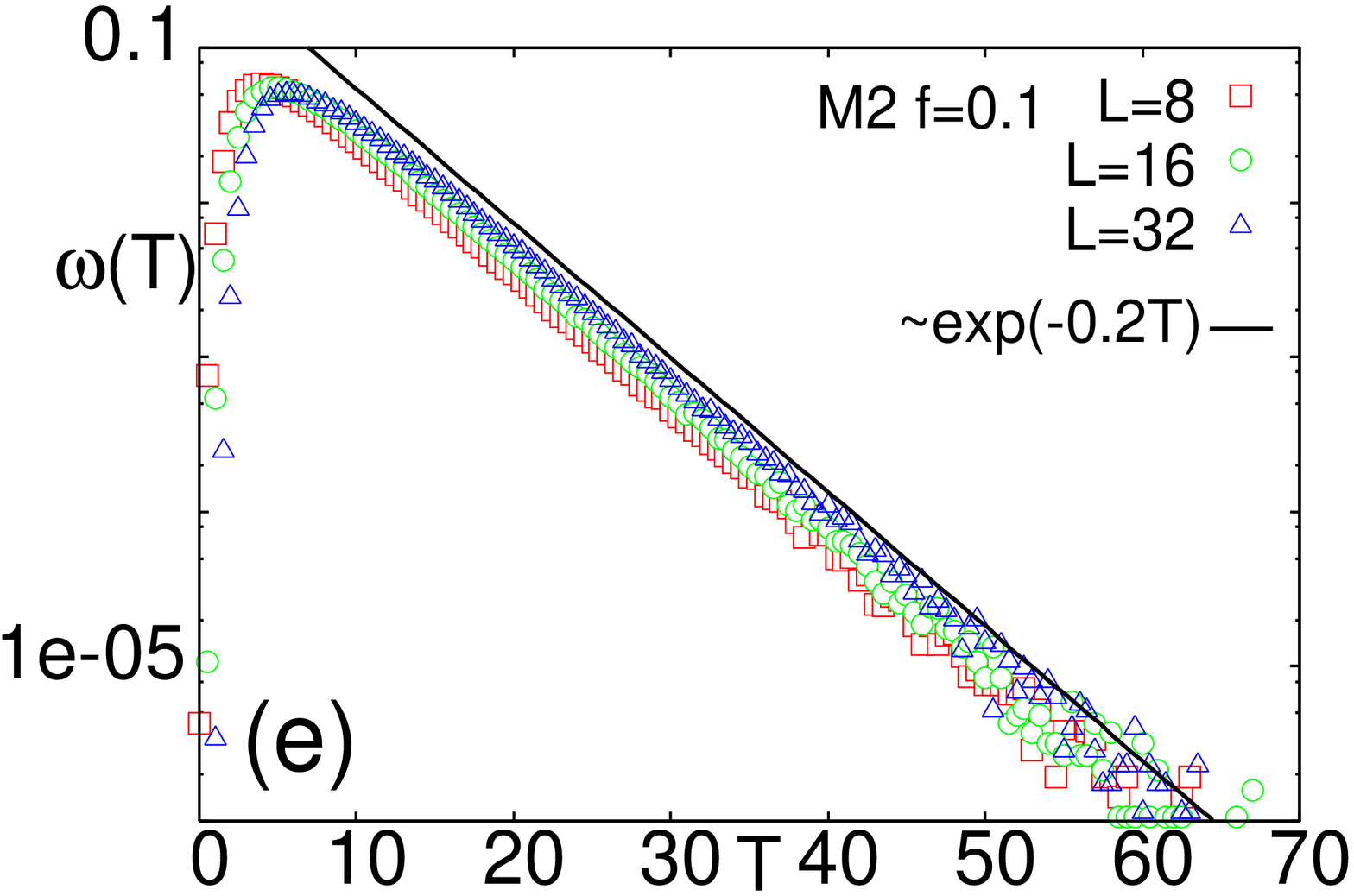}
\includegraphics[scale=0.25]{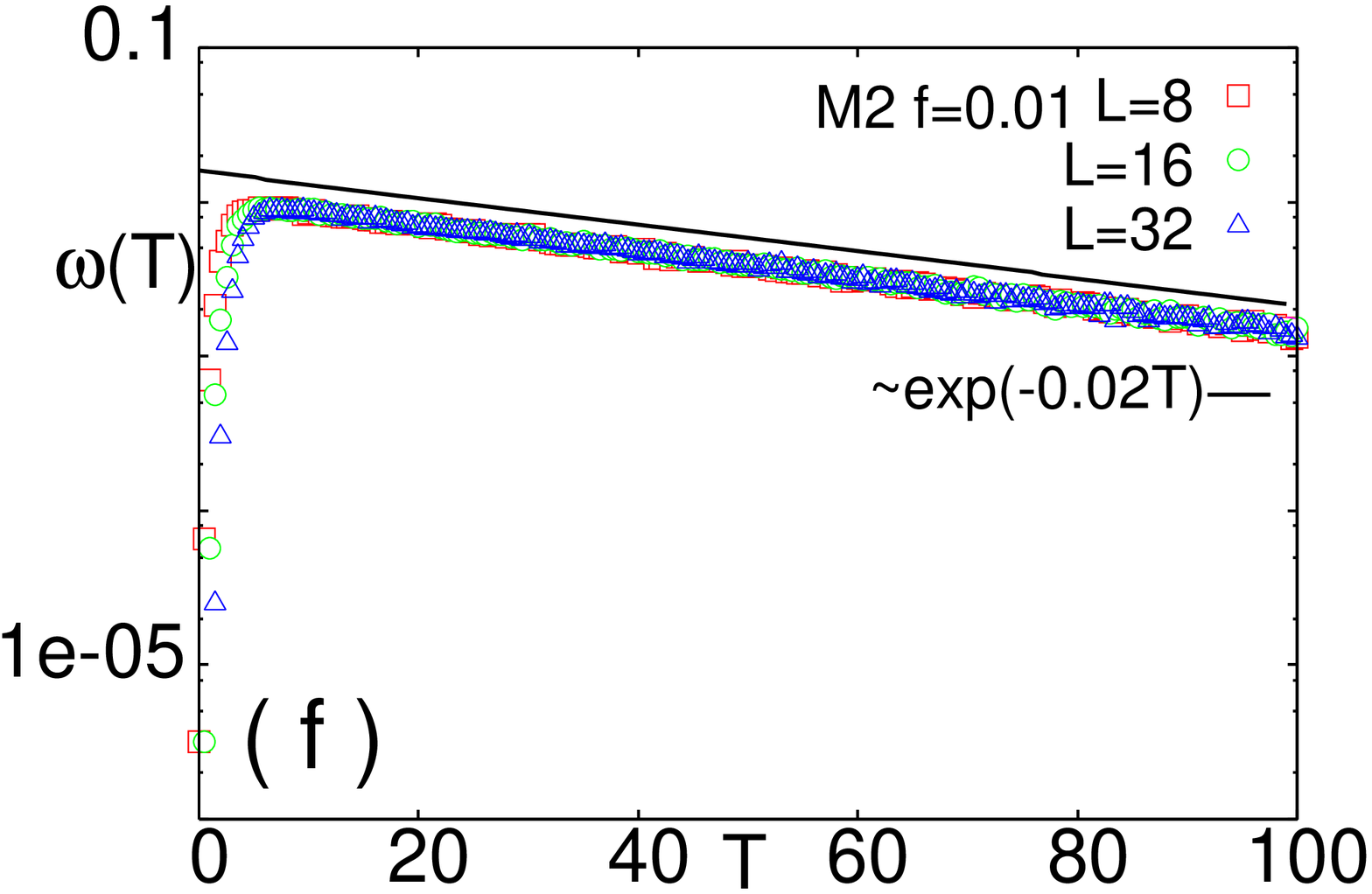}
\includegraphics[scale=0.25]{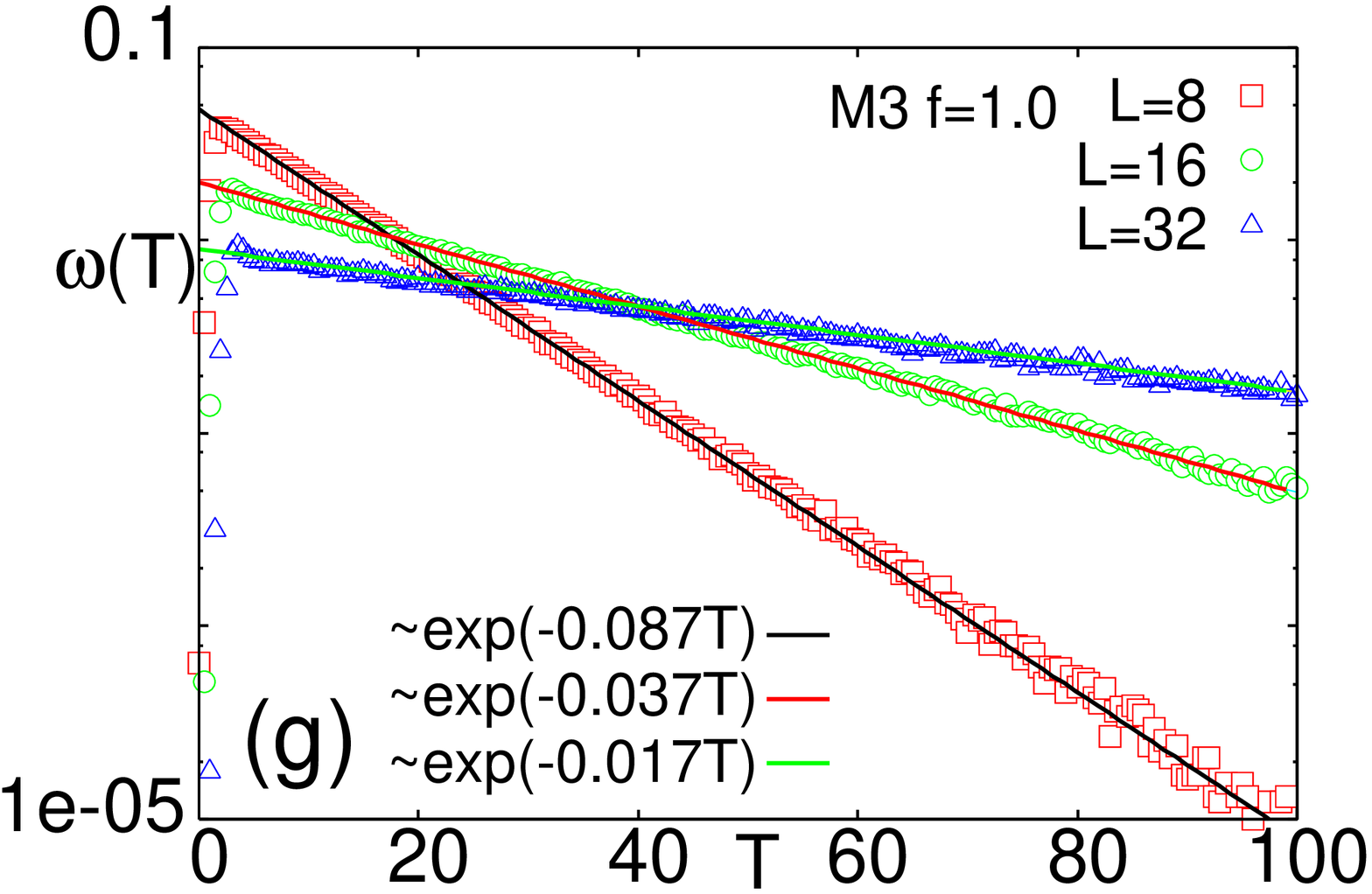}
\includegraphics[scale=0.25]{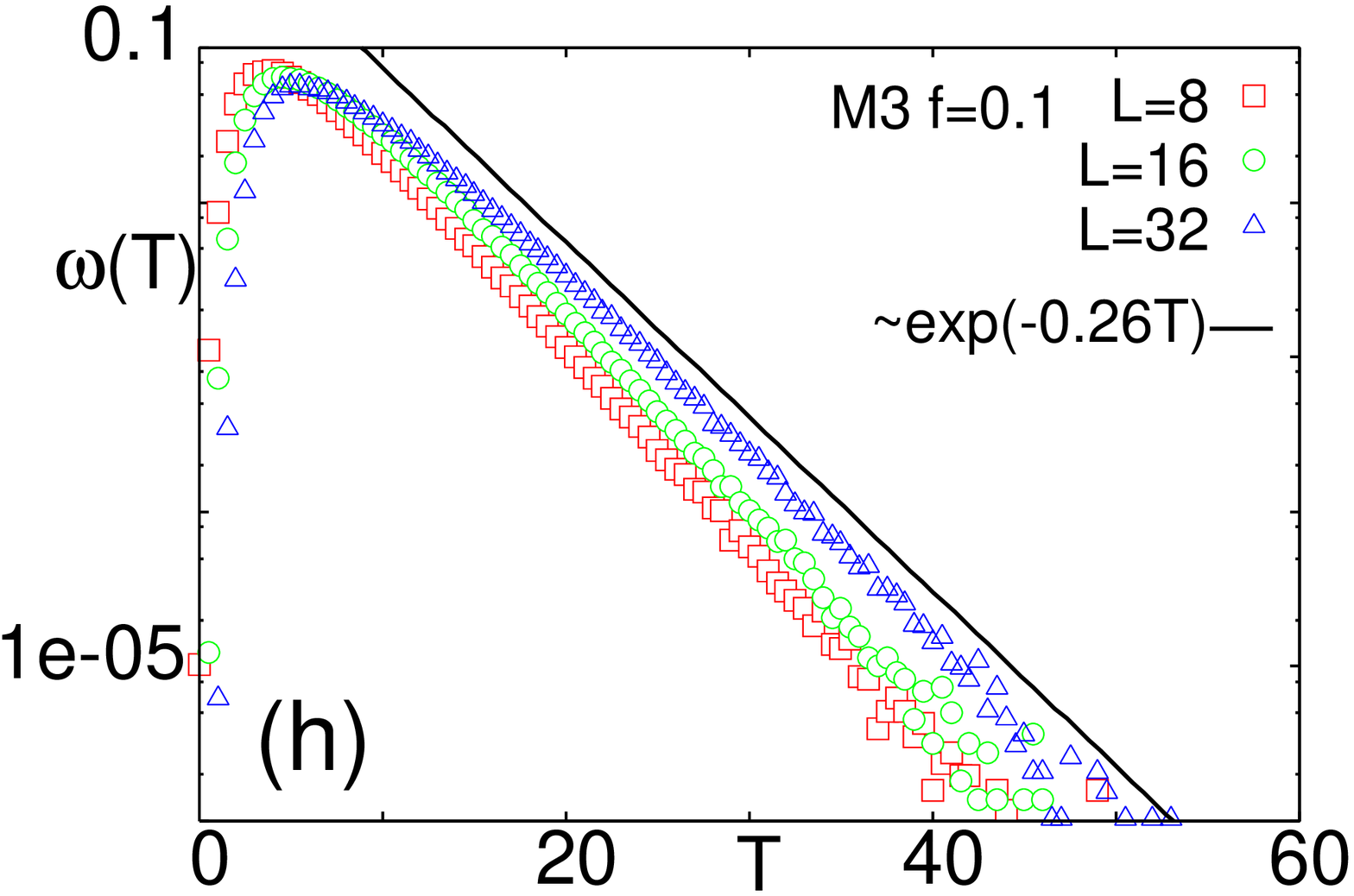}
\includegraphics[scale=0.25]{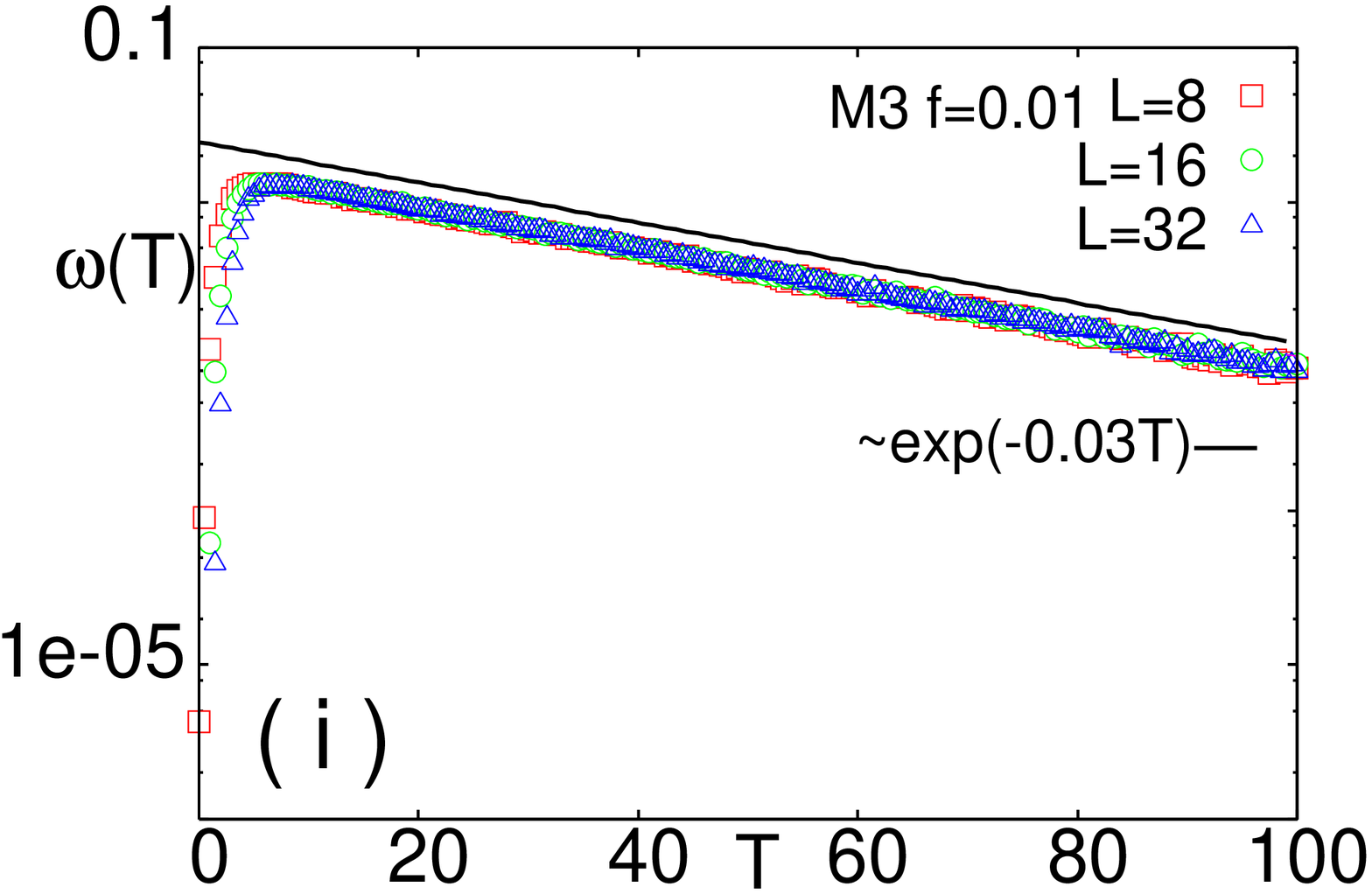}
\includegraphics[scale=0.25]{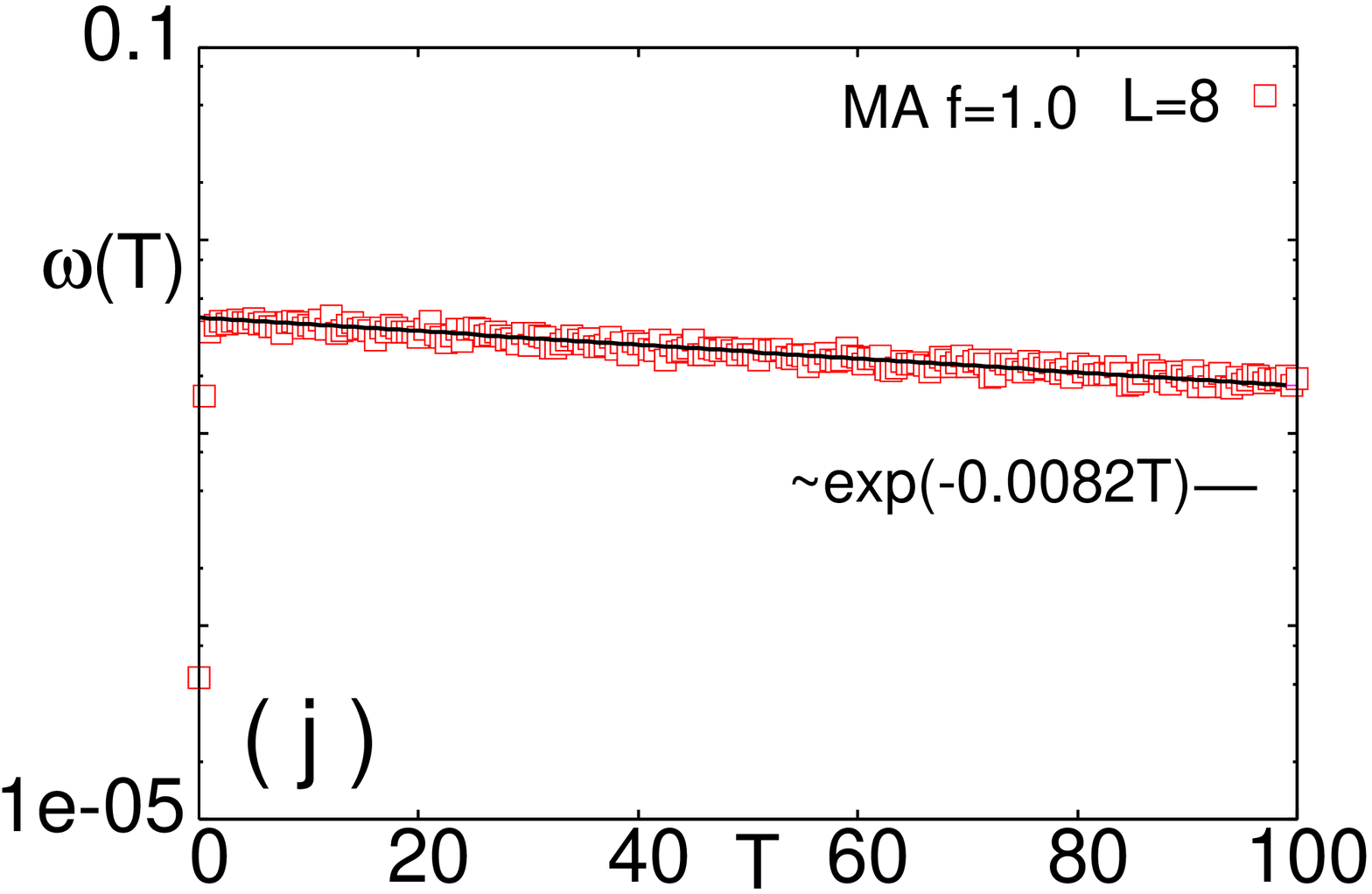}
\includegraphics[scale=0.25]{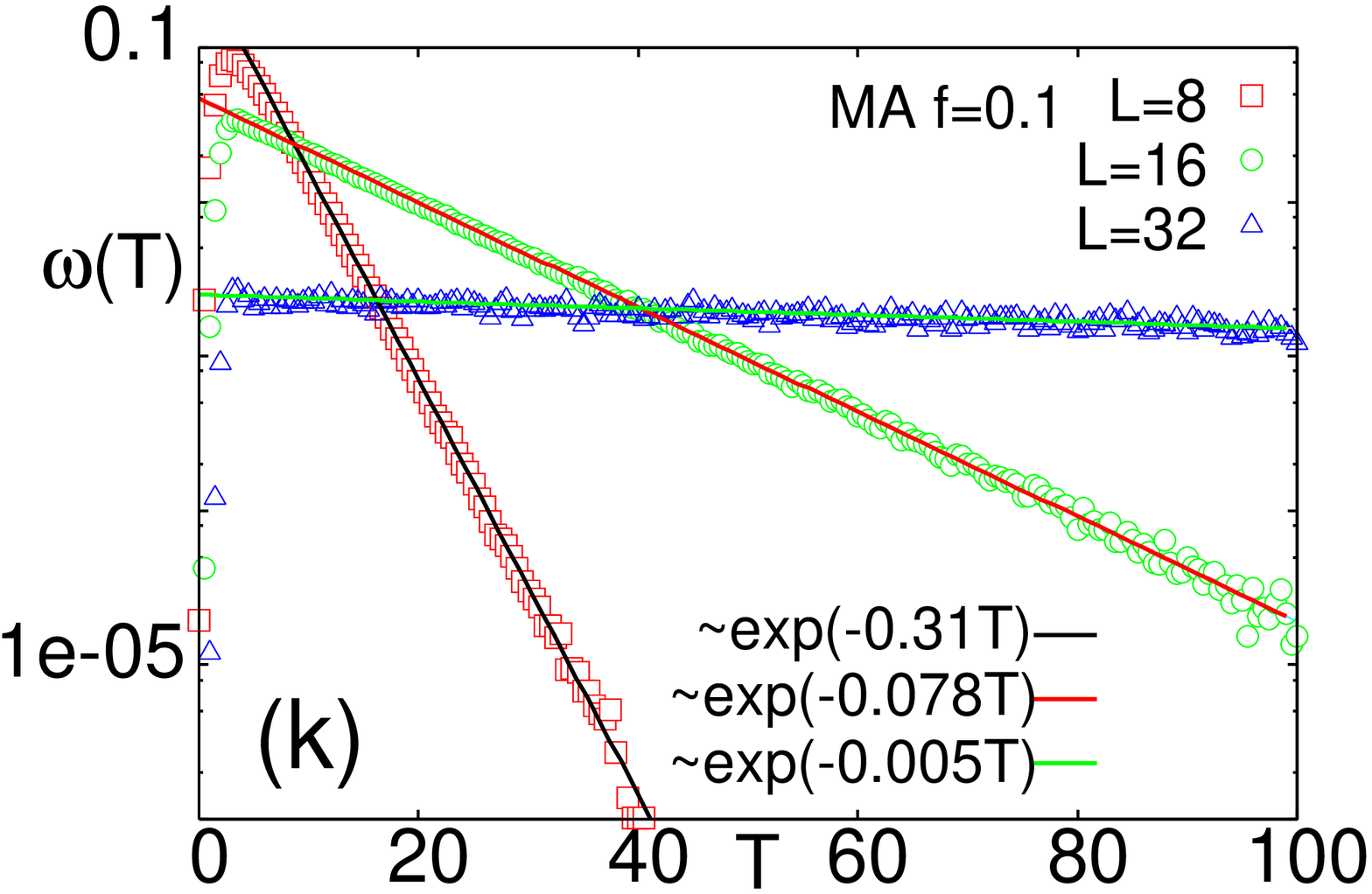}
\includegraphics[scale=0.25]{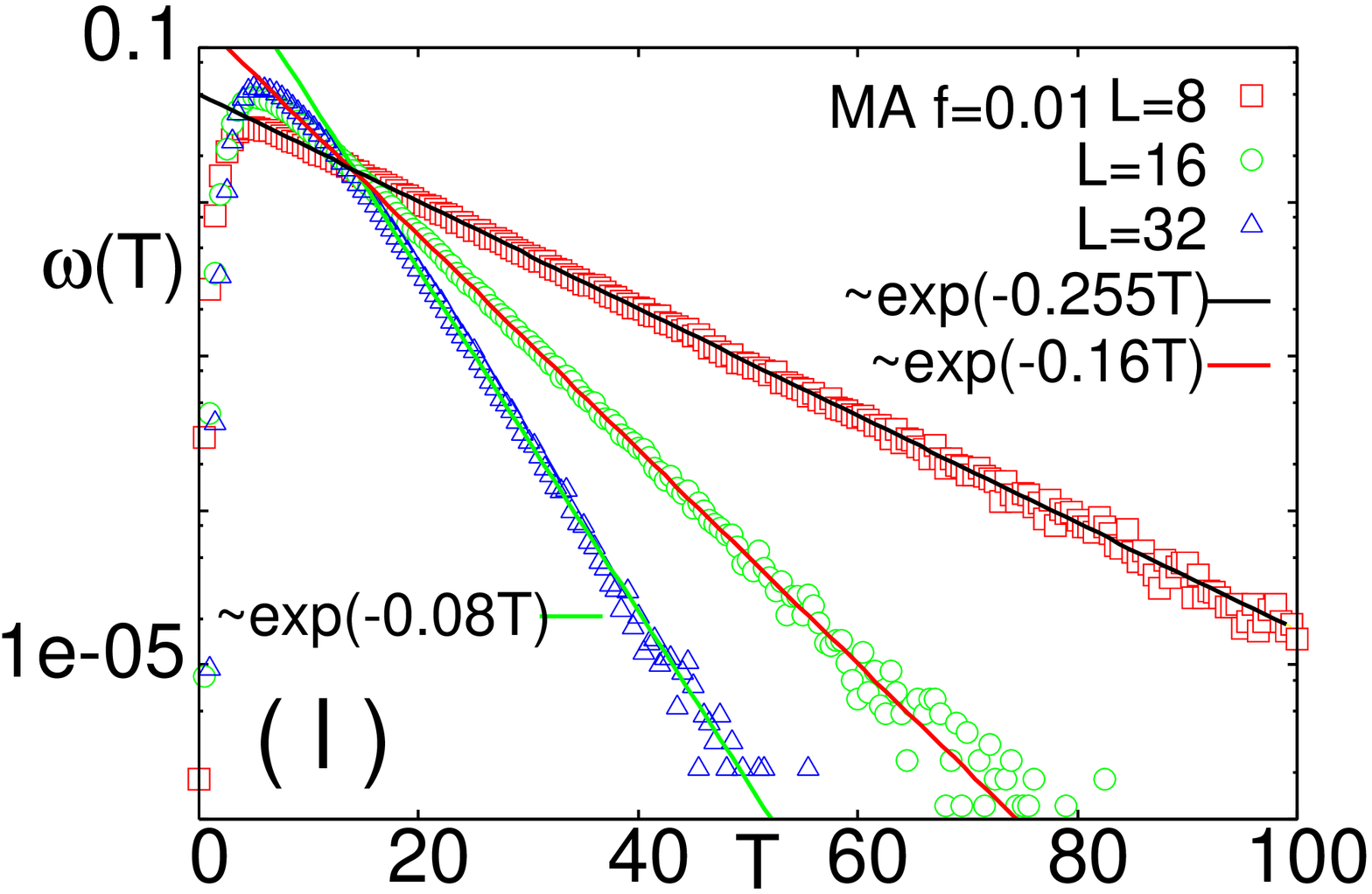}
\caption{The time-interval distributions of system-size earthquakes for 
 M1: $f=1.0$(a), $0.1$(b), $0.01$(c); M2: $f=1.0$(d), $0.1$(e), $0.01$(f);
 M3: $f=1.0$(g), $0.1$(h), $0.01$(i); and
 MA: $f=1.0$(j), $0.1$(k), $0.01$(l). The solid lines denote the fitted
 exponential or Weibull distributions.}
\end{center}
\end{figure}
For MA with $f=1.0$ (Fig.~6(j)), which is in the
Poisson phase, the distributions approximately 
obey exponential functions with long decay times. We omit the plots for
$L=16$ and $32$, because the
system-size events are so rare that very few occurred within the $2^{27}$ time
steps.
For M1, M2, and M3 with $f=0.01$ (Figs.~6(c), (f) and (i)), which are in the
two-state phase, 
the tail parts of the distributions are proportional to $e^{-fT},e^{-2fT}$ and
$e^{-3fT}$, respectively.
The decay rates $f, 2f$, and $3f$ are equal to the effective triggering
rates of M1, M2, and M3, respectively.
When the system is fully loaded, the occurrence rate of
system-size earthquakes is equal to the effective triggering rate, because the
fully loaded cluster includes all the trigger sites, which supports the
two-state transition picture.

Among the non-Poisson behaviors, the LDFs of M1 with $f=1.0$ (Fig.~5(a)),
and of M2 and M3 with $f=0.1$ (Fig.~5(e) and (h)), deviate greatly from
$\phi_P(z)$, as compared to the other non-Poisson cases.
In the time-interval distributions for non-Poisson phases (Fig.~6), we observe
prominent peaks and steep decays at large $T$. This can be
regarded as a characteristic of periodicity.
The tail parts of the distributions all decay exponentially, except for
M1 with $f=1.0$.

Figure~6(a) shows $\omega(T)$ for M1 with $f=1.0$, where the solid curves denote
functions proportional to 
the Weibull distributions $\sim(\y T)^{\g-1}\exp(-(\y T)^{\g})$ with 
$\g=1.45$. $\y$ is determined in each case by being fitted to the tail of the distribution.
The Weibull distribution is a good approximation to the tail of the time-interval distribution for M1 with $f=1.0$.
The Weibull distribution can also be found in the time-interval distribution
of real seismic activities and simulations \cite{Run,Has,Aki,TurT}.
The cause of the Weibull distribution in our results may be related
to the long-term memory.
Bunde et.al. suggested that a long-term memory leads to a stretched
exponential distribution of the time-interval distribution \cite{Bun2}.
The probability of large $T$ for $\g>1$ is smaller than that for $\g=1$
(simple exponential),
suggestive of the existence of some mechanism that shortens $\g>1$.
For M1 with large f, small earthquakes frequently occur, and these
release loads near the trigger site. Loads near the edge
opposite the trigger site tend to exist, and they are released only
by large earthquakes.
This mechanism cannot occur in M2, M3, or MA.
In M2, if a trigger also occurs on the opposite edge, 
this releases the existing loads and
makes the time intervals between the system-size earthquakes longer than they would be in
M1.

We observe exponential tails in 
the time-interval distributions for some cases of the
non-Poisson phase, contrary to the claim of Bunde
et.al. \cite{Bun,Bun2}.
This happens because non-Poisson behavior can also be caused by
periodic events.
The peak in the time-interval distribution clearly indicates that there is a 
periodicity in the occurrence of the system-size earthquakes.
Both the tail and the peak of the
time-interval distribution contribute to the
behavior of the LDF for the frequency of system-size earthquakes.

From the above observations, the non-Poisson behavior of system-size
earthquakes is mainly caused by the periodicity of their occurrence.
This picture is consistent with the hypothesis of the origin of the deep
trough in the LDF.
A deviation from exponential decay is observed in M1 with $f=1.0$,
although simple exponential decays are observed in the other models with the
parameters adopted in this study.
The simple exponential decay is a natural consequence because the simulation
is based on Poisson processes.
However, the exponential decay in the time-interval distribution may not
be caused by the Poisson processes in the model, since we observe
exponential decays in both the Poisson and the
non-Poisson phase.

\section{Discussion and Conclusions}

\subsection{The size-frequency distribution and the LDF of system-size
  earthquakes}

For the LDFs for MA with $f=0.1$ (Fig.~5(k)) and $0.01$ (Fig.~5(l)), which
belong to N(P), the mean frequency $x_m$ shows that the scaling failed.
For MA with $f=1.0$: (i) the LDF shows the Poisson phase (Fig.~5(j)); (ii)
there is no peak for the system-size earthquakes in the size-frequency
distribution (Fig.~3 (d)); and (iii) the mean frequency of the system-size
earthquakes can be approximated by an exponential function of $L$ (Fig.~4(b)).
The characteristics (ii) and (iii) hold for $f=0.1$ and $0.01$ when the system
size is sufficiently large; the details are shown below.

Figure~7 shows the size-frequency distributions of earthquakes for MA with
$f=1.0, 0.1$, and $0.01$.
\begin{figure}
\begin{center}
\includegraphics[scale=0.25]{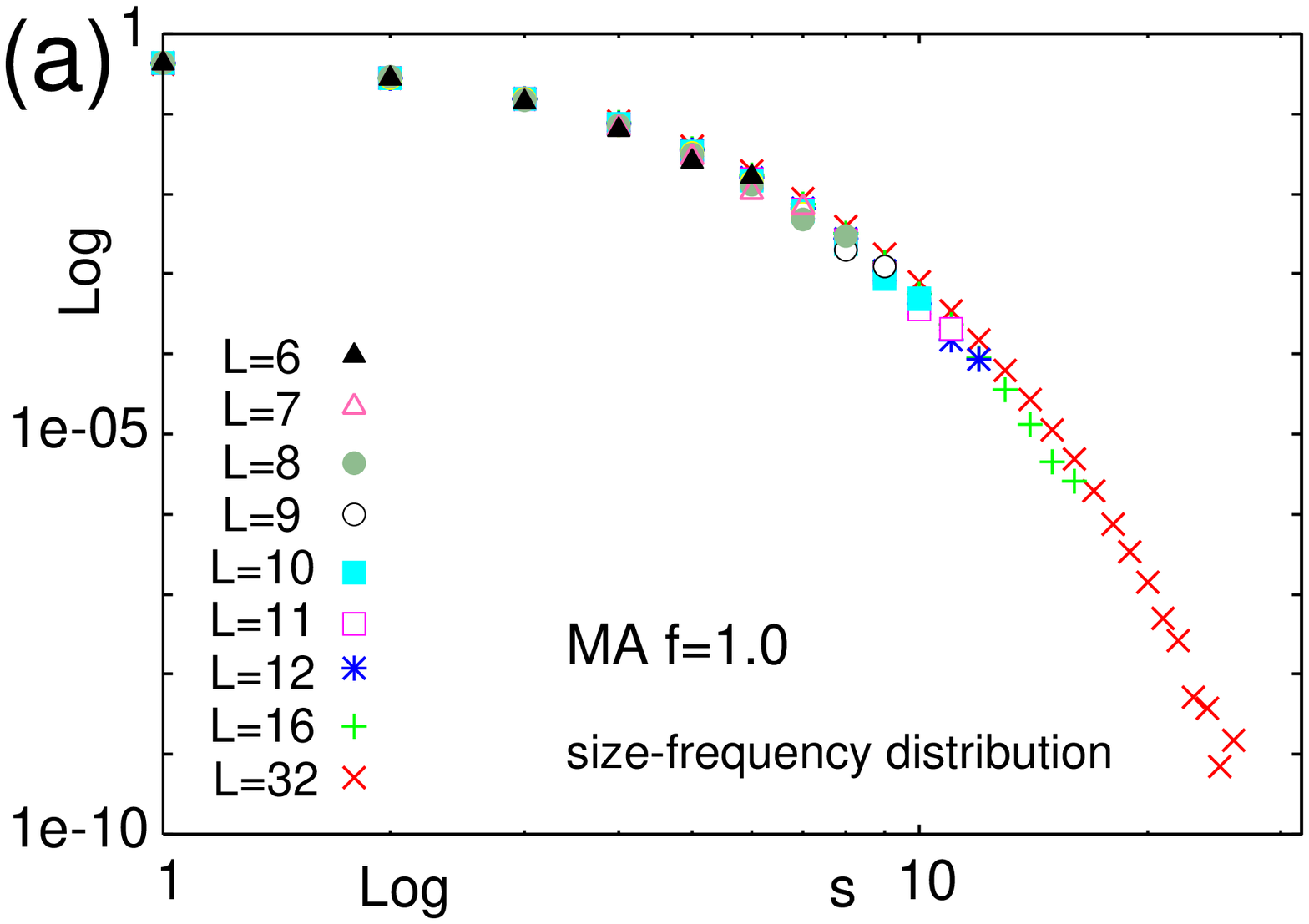}
\includegraphics[scale=0.25]{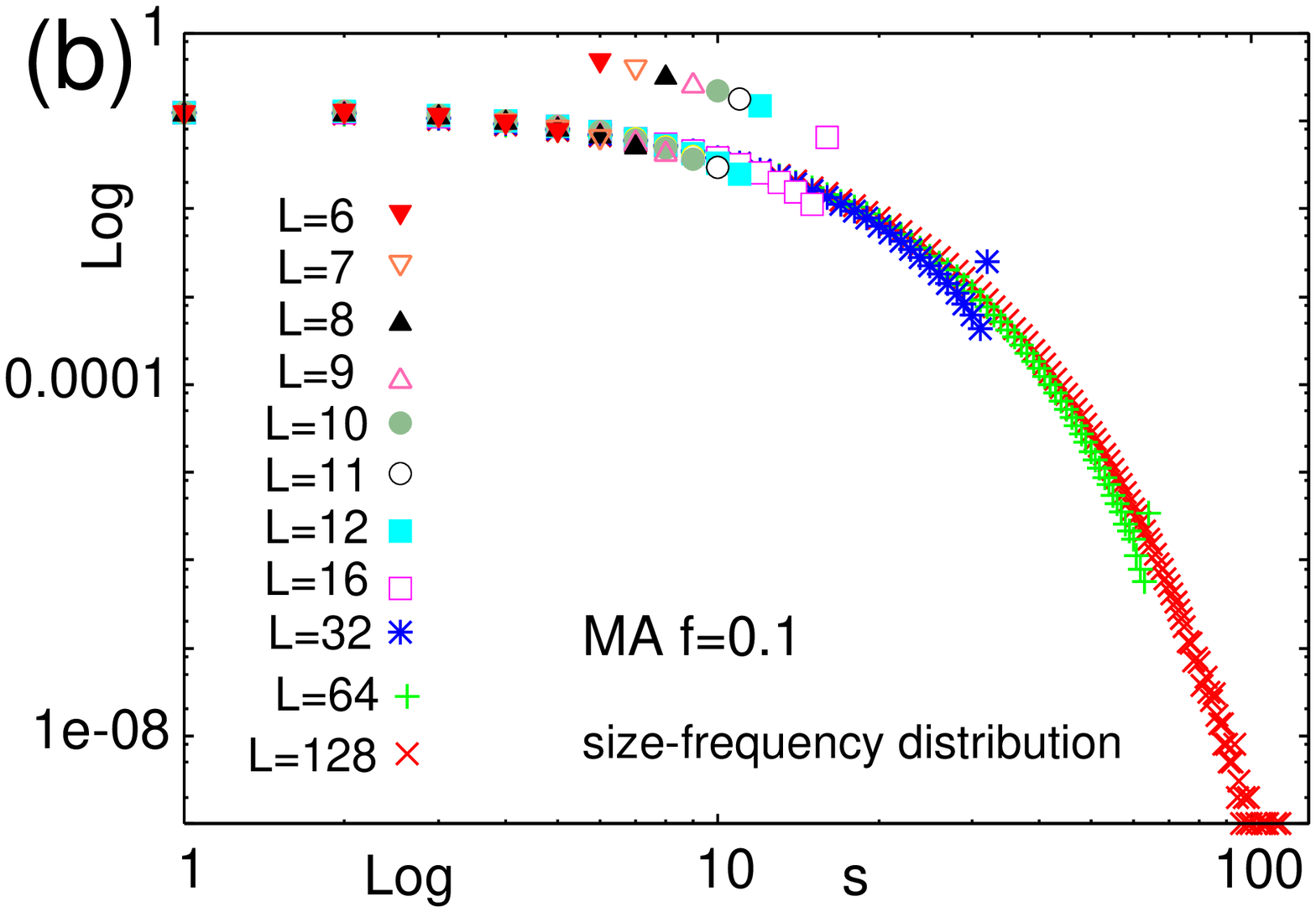}
\includegraphics[scale=0.25]{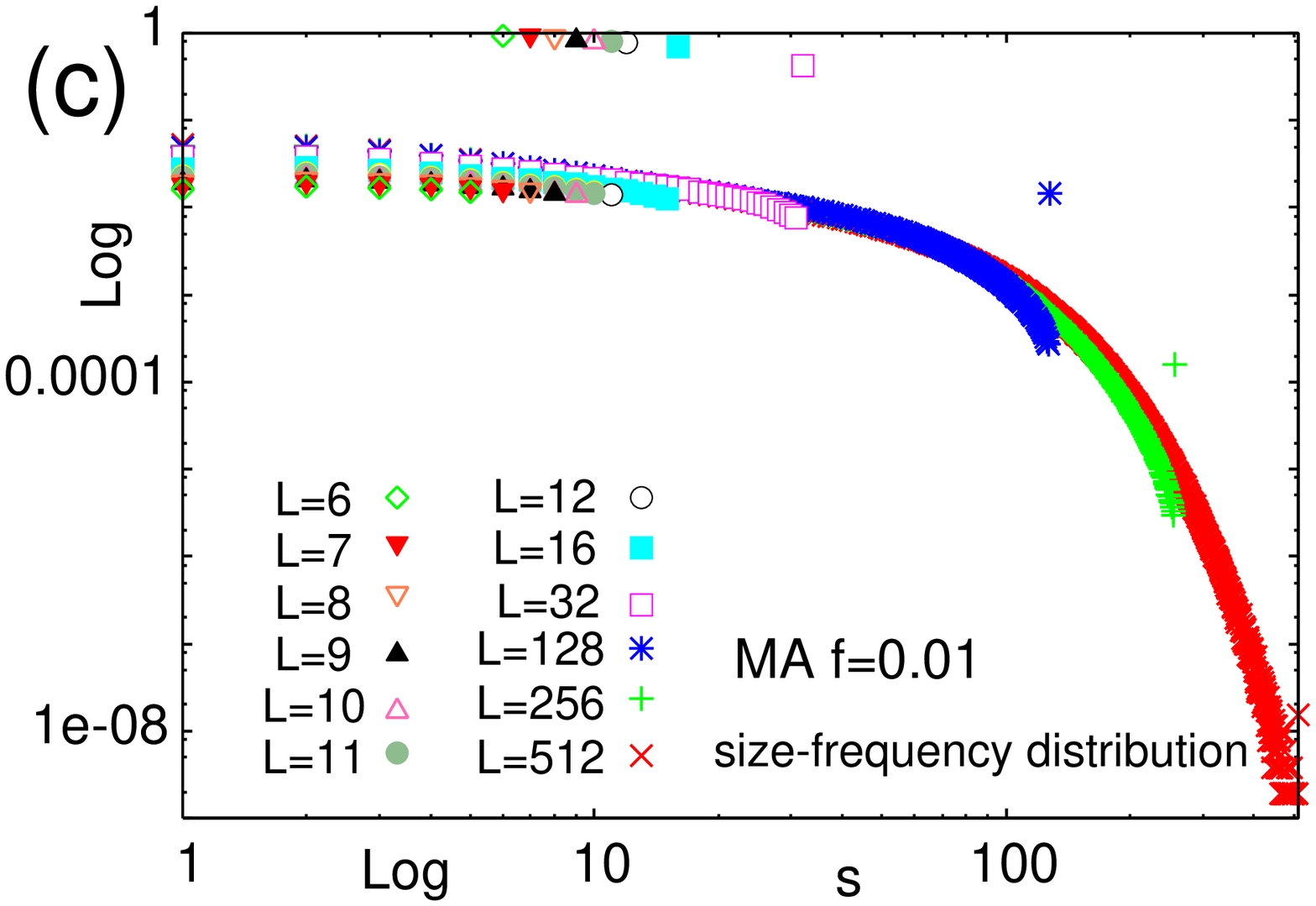}
\caption{Size-frequency distributions of MA for different $L$ with (a) $f=1.0$, (b) $f=0.1$,
and (c) $f=0.01$.}
\end{center}
\end{figure}
The distributions for MA with $f=1.0$ do not show peaks at
$s=L$, while those for $f=0.1$ do show peaks at $s=L$.
For MA with $f=0.1$ and $0.01$, the peak frequency at $s=L$ is clear
when $L$ is small, but it becomes unclear as $L$ increases.
This suggests that characteristic (ii) is satisfied for large $L$.

Figure~4(b) shows that the mean frequency of the system-size earthquakes behaves asymptotically as an exponential function of
$L$ for MA with large $L$, indicating that characteristic (iii) holds
for large $L$.
Thus we confirm that characteristics (ii) and (iii) are satisfied for
large $L$ and MA with $f=0.1$ and $0.01$, and we expect that the
LDF can be approximated by the Poisson LDF for these cases.

In the MA model with large $L$ and $f$, earthquakes are
frequently triggered before the system is fully loaded.
This results in the occurrence of many small earthquakes and few system-size
earthquakes, leading to characteristics (ii) and (iii). 
The configuration of a fully loaded system appears rarely and randomly, 
and therefore the sequence of system-size earthquakes 
approximately follows a Poisson process.
Thus we expect that the LDFs of system-size earthquakes in the MA model
may be categorized as Poisson phase for large $L$.
This is supported by simulation results, at least for small $z$.
Figure~8 shows $-\frac{1}{x_mt}\log{P(z)}$, which approximates the LDF for
large $t$, calculated by the normal Monte Carlo for MA with
$L=48,64,96$ and $128$ and $f=0.01$.
The simulation is conducted $2^{14}$ time steps for each run.
As $L$ increases, the data approach the LDF of the Poisson process
$\phi_P(z)$.
Note that this expectation has not been proven theoretically, which is left for
a future study.
\begin{figure}
\begin{center}
\includegraphics[scale=0.4]{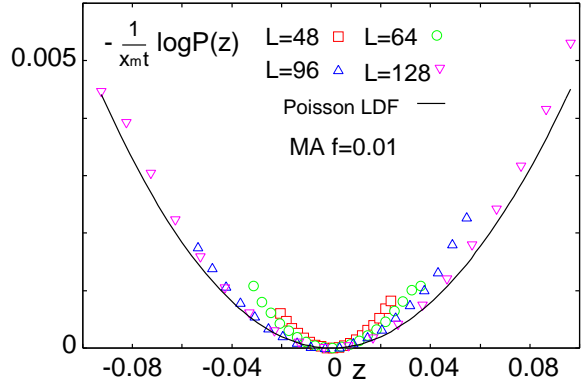}
\caption{$-\frac{1}{x_mt}\log{P(z)}$ calculated by the normal Monte Carlo for MA with
$L=48,64,96$, and $128$ and $f=0.01$.}
\end{center}
\end{figure}

So far, for the MA model, we have discussed the relations among the LDFs for the frequency of
system-size earthquakes, the size-frequency distribution of earthquakes,
and the dependence of the mean frequency of system-size
earthquakes on the system size, and we have found that the characteristic time
sequence of system-size earthquakes is related to the size-frequency
distribution.
For the non-Poisson behaviors of M1, M2, and M3, we find vague
relations between the size-frequency distributions and
the LDFs for the frequency of system-size earthquakes.
For M3 with $f=1.0$ and increasing $L$, the LDF approaches the LDF of
the Poisson process (Fig.~5(g)), which may be similar to the
N(P) phase (Fig.~5(k)).
However, in this case, neither characteristic (ii) nor (iii) is
satisfied, suggesting that the behavior is different from that of the MA
model.
We also calculated the time-interval distributions in order to gain more
information about the relations between the phases.
The non-Poisson behavior is related to the peak and a steep decay in
the time-interval distribution, as shown in Fig.~6.
The peak and the steep decay correspond to the deep trough in the LDF for the
frequency.
In the cases for which the size-frequency distribution of the system-size earthquakes demonstrates a clear periodicity (M1 with
$f=1.0$, M2 with $f=0.1$, and M3 with $f=0.1$), $N(L)/N_A$ 
is not very close to
$1$, as shown in Fig.~3, indicating that the system-size earthquakes do
not dominate the small earthquakes.
The characteristics of the non-Poisson phase that are observed in the
size-frequency distributions, the LDFs for frequency, and the time-interval distributions may be explained by the periodicity of the
system-size earthquakes.
Thus in some cases, the non-Poisson behaviors in the LDFs correspond to the phases of the
size-frequency distributions.

\subsection{Conclusion}

The size-frequency distributions and the LDFs for the frequencies of system-size earthquakes for four kinds of 1D forest-fire models were
calculated numerically.
The LDFs for the frequency of the system-size earthquakes mostly deviate
from the LDF of the Poisson process.
We classified the behaviors of the size-frequency distributions into three
types: supercritical, critical, and subcritical.
We also classified the behaviors of the LDFs for the frequency of the
system-size earthquakes into three types: Poisson, non-Poisson, and
two-state.
The Poisson phase of the LDF is related to the subcritical
phase, where the peak for system-size earthquakes is not clear in the
size-frequency distribution.
This relation has yet to be confirmed by the calculation of the LDF for
large $L$, where the mean frequency of system-size earthquakes
exponentially decreases with an increase in system size.
The calculation of the LDF for such a large $L$ is difficult at
present.

For real seismic activities, the statistical properties of the
frequency of large earthquakes of magnitudes greater than 7.0 were studied
in Daub et.al. \cite{Dau}, and it was found that the sequence of large earthquakes obeys
the Poisson process for magnitudes over 7.3.
Although the 1D forest-fire models are too simple to characterize the
complex properties of real earthquakes, 
our findings of the correspondence between the subcritical behavior of
the size-frequency distribution and the good approximation by the
Poisson LDF for the MA model may relate the subcritical magnitude-frequency
distribution of real earthquakes to the random occurrences of system-size earthquakes.
Whether the system-size earthquakes occur at random, obeying the Poisson
process, or with some regular periodicity, depends on the properties of the
heterogeneity of faults.

The present study focused on earthquakes.
Deviation from the Poisson process appears in many scientific
contexts, such as bunching in the process of 
photon counting and spike trains of neurons.
The LDF approach can also be applied to those phenomena, and
we hope this study furthers the understanding of the underlying physics of
systems described by point processes.

\section{Acknowledgements}
This research was supported by the MEXT project ``Evaluation and disaster
prevention research for the coming Tokai, Tonankai and Nankai
earthquakes''.
T. M. is supported by the Aihara Project, the FIRST program from the
 JSPS, initiated by the CSTP.
The numerical calculations in this study were partly carried out at the 
 YITP at Kyoto University.

\appendix

\section{Exact forms of $W(CC')$}

$W(CC')$ is written as
\begin{eqnarray}
\label{wcc}
 W(CC') &=& p \sum_{j=1}^L
  [\cdots\d_{\t_{j-1}\t_{j-1}'}(1-\t_j')\t_j
  \d_{\t_{j+1}\t_{j+1}'}\cdots] +W_f(CC'),
\end{eqnarray}
where $W_f(CC')$ is the trigger term that depends on the model, and
$\d_{xx'}$ is the Kronecker delta.
The parts that have a suffix less than $1$ or greater than $L$ are replaced by
unity (e.g., $\d_{\t_0\t_0'}$ is replaced by $1$.).
The exact forms of $W_f(CC')$ are written as
\begin{equation}
  W_f(CC')=
   f\sum_{j=1}^{L}[\t_1'(1-\t_1)\cdots\t_j'(1-\t_j)
   (1-\t_{j+1}')(1-\t_{j+1}) \d_{\t_{j+2}\t_{j+2}'}
   \cdots ]
\end{equation}
for M1;
\begin{eqnarray}
& & W_f(CC')= W_2 =
  f\sum_{j=1}^{L}[\t_1'(1-\t_1)\cdots\t_j'(1-\t_j)
  (1-\t_{j+1}')(1-\t_{j+1})\d_{\t_{j+2}\t_{j+2}'} \cdots]
   \nonumber \\ & & +
  f\sum_{j=1}^{L}[\t_{L-j}'(1-\t_{L-j}) \cdots
  \t_L'(1-\t_L)(1-\t_{L-j-1}')(1-\t_{L-j-1})
  \d_{\t_{L-j-2}\t_{L-j-2}'} \cdots ]  \nonumber \\
\end{eqnarray}
for M2;
\begin{eqnarray}
 W_f(CC') &=& W_2+ f\sum_{j=1}^{m}\sum_{k=m}^{L}
  [\cdots\d_{\t_{j-2},\t_{j-2}'}
  (1-\t_{j-1}')(1-\t_{j-1})\t_j'(1-\t_{j})\cdots
  \nonumber \\ & &
  \t_{k}'(1-\t_{k})(1-\t_{k+1}')(1-\t_{k+1})
  \d_{\t_{k+2}\t_{k+2}'}\cdots ] 
\end{eqnarray}
for M3; and
\begin{eqnarray}
W_f(CC') &=& W_2+ f\sum_{m=2}^{L-1} \sum_{j=1}^{m} \sum_{k=m}^{L}
  [\cdots\d_{\t_{j-2},\t_{j-2}'}
  (1-\t_{j-1}')(1-\t_{j-1})\t_j'(1-\t_{j})\cdots
  \nonumber \\ & &
  \t_{k}'(1-\t_{k})(1-\t_{k+1}')(1-\t_{k+1})
  \d_{\t_{k+2}\t_{k+2}'}\cdots]
\end{eqnarray}
for MA.

\end{document}